\documentclass[11pt,onecolumn,amssymb,nofootinbib,floatfix]{revtex4}
\usepackage{amsmath, amsthm, amscd, amssymb}
\usepackage{graphicx, braket}
\usepackage{bm}
\usepackage{bbm}
\usepackage{epstopdf}
\begin{document}

\title{\bf Spherically symmetric vacuum solutions arising from trace dynamics modifications to
gravitation}

\author{Stephen L. Adler}
\email{adler@ias.edu} \affiliation{Institute for Advanced Study,
Einstein Drive, Princeton, NJ 08540, USA.}

\author{Fethi M. Ramazano\v glu}
\email{framazan@Princeton.edu} \affiliation{Department of Physics, Princeton University, Princeton, NJ 08544, USA.}

\begin{abstract}
We derive the equations governing static, spherically symmetric vacuum solutions to the
Einstein equations, as modified by the frame-dependent effective action (derived from trace dynamics) that gives an alternative explanation of the
origin of ``dark energy''.
We give analytic and numerical results for the solutions of these equations,
first in polar coordinates, and then in isotropic coordinates.  General features
of the static case are that (i) there is no horizon, since $g_{00}$ is non-vanishing for finite
values of the polar radius, and only vanishes (in isotropic coordinates) at the internal singularity, (ii) the Ricci scalar $R$ vanishes
identically, and (iii) there is a physical singularity at cosmological distances.  The large
distance singularity may be an artifact of the static restriction, since we find that the behavior at large distances is altered in a time-dependent solution using the McVittie Ansatz.

\end{abstract}

\maketitle

\section{Introduction}
\subsection{Background and aims}
In a recent paper \cite{adlerpaper}, one of us (SLA)  analyzed the consequences of introducing gravitation into
the framework of trace dynamics pre-quantum mechanics \cite{adlerbook}, assuming the metric to be described
as usual by a classical field.   The focus of \cite{adlerpaper} was on deriving an induced effective action describing
trace dynamics modifications to gravitation, applying it to the cosmological Robertson-Walker metric, and beginning
an application to the spherically symmetric case.  The purpose of the present paper is to continue the latter,
with a detailed investigation of modifications to spherically symmetric vacuum solutions.

Before proceeding with this, however, we give a brief introduction to the basic ideas of trace dynamics pre-quantum
mechanics.   A book-length exposition of trace dynamics is given in [2], with a
synopsis as section 2
of [1]; readers wanting more detail than given here can refer to these references.
The fundamental idea
of trace dynamics is to set up a pre-quantum theory, based on
non-commuting matrix-
or operator- valued dynamical variables which have no a priori
commutation relations; for example,
canonical coordinates $q_r$ at different spatial points are not assumed
to commute.
The Lagrangian for this system is taken as the trace of an operator
Lagrangian constructed
as a polynomial in the $q_r$ and their time derivatives, and using just
cyclic permutation
under the trace, one finds that the Euler Lagrange equations for the
trace Lagrangian
are operator equations of motion, without invoking canonical
quantization.  This system
is more general than the usual quantum field theory, and the rest of the
trace dynamics
program shows that with certain approximations (basically a decoupling
of high energy
from low energy degrees of freedom), quantum field theory with unitary
state evolution
emerges as the thermodynamics  of the underlying trace dynamics system.
In trace
dynamics, state vector reduction in measurement is a physical process, and is  the only
low energy
non-local and non-unitary remnant of the underlying operator dynamics.

Taking thermodynamics averages requires a canonical ensemble, which in
trace dynamics
is constructed as usual as the exponential of a sum of generic conserved
quantities with
Lagrange multiplier coefficients.  Thus, the canonical ensemble has the
form
\begin{equation} \label{eq:rhodef}
d\mu\rho=\frac{d\mu\exp[-\tau {\bf H} + ...] }{\int d\mu \exp[-\tau {\bf
H}+...]}~~~.
\end{equation}
Here $d\mu$ is the integration measure over the underlying phase space of
pre-quantum operator variables, which is conserved by the dynamics,
${\bf H}$ is
the trace Hamiltonian, which is also conserved,  $\tau$ is the Lagrange
multiplier
for the trace Hamiltonian, and the elipsis ... denotes further
conserved quantities that are Lorentz invariant when the trace
Lagrangian is assumed
Lorentz invariant.  The main point that concerns us here is that ${\bf
H}$ is not
Lorentz invariant; it is rotationally invariant, but picks out a
preferred frame specified
by a time-like unit four vector $\eta=(1,0,0,0)$.  Thus before
decoupling of terms
proportional to $\tau$, averages over the canonical ensemble will have a
frame
dependence.  This can be interpreted as a reflection of the rest frame
of the ``fluid''
consisting of the pre-quantum degrees of freedom.   Since the trace
Hamiltonian, and
the entire formalism, are rotationally invariant, and more generally are invariant
under three-space general coordinate transformations, any preferred frame
effects arising from
the canonical ensemble share these invariances.

In the discussion of the book [2], gravity was not included.  In the
paper [1] , gravity
was included in the form of a classical background metric $g_{\mu\nu}$, and a trace
dynamics-induced
correction to the usual Einstein action was defined as the average of
the trace dynamics
pre-quantum variable action over the canonical ensemble.  The structure
of this correction was analyzed under the assumption that the
pre-quantum matter fields are massless, and so
admit a Weyl scaling invariance when the metric is rescaled.  This, together with three space rotational and general coordinate invariance,
implies that the induced correction to the Einstein action for has the
following form for non-rotating
metrics, to zeroth order in the derivatives of the metric,
\begin{equation}\label{eq:action1}
\Delta S_{\rm g}= A_0  \int d^4x (^{(4)}g)^{1/2}(g_{00})^{-2}  ~~~,
\end{equation}
with $A_0$ a constant and with $^{(4)}g=-{\rm det} g_{\mu\nu}$.

A salient feature of this effective action is that it picks a preferred frame, and thus violates Lorentz invariance.
The traditional statement of the principle of relativity is that all inertial frames are physically equivalent,
and that there is no way to measure the absolute velocity of an inertial observer.  We now know, since the discovery of the cosmic microwave
background (CMB) radiation, that this statement needs modification.  The CMB provides a reference inertial frame, and by measuring the dipole
component of the angular variation of the CMB, an observer with a radiometer can infer her absolute velocity with respect to this frame.  Such measurements \cite{planck}
show that the solar system is moving with a velocity of 369 km/s relative to the CMB. Empirically, we know that the CMB is highly isotropic, to an
accuracy of one part in $10^5$.  Thus, it is natural to identify the preferred frame picked out by the trace dynamics canonical ensemble, and by
the action of Eq. \eqref{eq:action1}, with the CMB rest frame, since identification with any frame moving with a finite velocity with respect to the
CMB rest frame would violate rotational symmetry and spatial isotropy.  Calculations using the effective action of Eq. \eqref{eq:action1} must be
carried out in the CMB rest frame, with predictions for  frames moving with a finite velocity with respect to the CMB frame obtained by applying an appropriate boost to the results of the CMB rest frame calculations.  None of this implies that the physical laws underlying trace dynamics are
Lorentz violating -- they are not -- but only reflects the rest frame of
the underlying pre-quantum ``fluid'', the dynamics of which is averaged over by the trace dynamics canonical ensemble.  Our assertion, motivated by
rotational invariance, is that this rest frame is the same as the rest frame of the CMB, as well as the rest frame of the  background galaxies.

There has been an extended discussion in the literature  of possible Lorentz violating effects,
and experimental bounds on them, but to our knowledge these do not take the form of a Lorentz violating effective action of the form of Eq. \eqref{eq:action1}.    Thus, one motivation of this paper is to study further implications of a new form of preferred
frame effect, based on the two assumptions which are used in \cite{adlerpaper} but which are more general than the trace dynamics program: (1) the
preferred frame effective action, as a function of the metric, is three space general coordinate invariant  (but not four space general coordinate invariant),  and (2) the preferred
frame effective action is invariant under Weyl rescalings of the metric and matter fields.

The simplest place to begin studying the implications of adding the action of Eq. \eqref{eq:action1} to the usual Einstein--Hilbert action
is the Robertson-Walker (RW) cosmological
line element,
\begin{equation}\label{eq:rwline}
ds^2=dt^2 - a(t)^2\left[\frac{dr^2}{1-kr^2} + r^2 (d\theta^2 + \sin^2\theta d\phi^2)\right]~~~,
\end{equation}
corresponding to the metric components
\begin{equation}\label{eq:rwcomponents}
g_{00}=1~,~~g_{rr}=-a(t)^2/(1-k r^2)~,~~g_{\theta \theta}=-a(t)^2 r^2~,~~g_{\phi \phi}=-a(t)^2 r^2 \sin^2\theta~~~.
\end{equation}
Since $g_{00}$ is exactly one for the RW line element, the effective action of Eq. \eqref{eq:action1} reduces in this case to
\begin{equation}\label{eq:action11}
\Delta S_{\rm g}= A_0  \int d^4x (^{(4)}g)^{1/2}  ~~~,
\end{equation}
and so has {\it exactly} the form of a cosmological constant action.
 This offers the possibility of explaining
the ``dark energy'' that contributes the largest fraction of the closure density of the universe, not as
reflecting a ``bare'' cosmological constant $\Lambda_0$ (see Appendix A for notation) associated with Lorentz invariant
quantum fluctuations, but rather as a manifestation of a frame-dependent effective action produced by pre-quantum matter field fluctuations.  In this spirit, we shall
assume that $\Lambda_0=0$, and we shall choose $A_0$ in Eq. \eqref{eq:action1} to reproduce the observed
cosmological constant $\Lambda$ for a RW cosmology, which requires \cite{adlerpaper}
\begin{equation}\label{eq:lambda}
\Lambda = - 8 \pi G A_0~~~,
\end{equation}
with $G$ the gravitational constant (which will be set equal to unity in perturbation expansions and the presentation
of numerical results).

We can then rewrite $\Delta S$ of Eq. \eqref{eq:action1} as
\begin{equation}\label{eq:action2}
\Delta S_{\rm g}=-\frac{\Lambda}{8\pi G}\int d^4x (^{(4)}g)^{1/2}(g_{00})^{-2} ~~~,
\end{equation}
which is to be added to the standard Einstein-Hilbert action
\begin{equation}\label{eq:einsteinaction}
S_{\rm g} =  \frac{1}{16\pi G} \int d^4x (^{(4)}g)^{1/2} R~~~
\end{equation}
to give the total gravitational action
\begin{equation}\label{eq:totalaction}
S_{\rm total}=S_{\rm g}+\Delta S_{\rm g}~~~.
\end{equation}
Varying this action with respect to the spatial metric components $g_{ij}$ gives the
modified  equations of motion for the spatial components $G_{ij}$ of the Einstein tensor;
as discussed in \cite{adlerpaper} and as elaborated on below, the equations of motion
of the remaining components of the Einstein tensor can be inferred by conserving extension
using the Bianchi identities.

By construction, for the RW metric the modified action of Eq. \eqref{eq:totalaction} gives the
usual equations of the standard cosmological model.  But for other metrics, it implies
different physics than is implied by  the usual interpretation of ``dark energy'' as a consequence
of a ``bare'' cosmological term, and this offers the possibility of distinguishing between
the two interpretations of ``dark energy''.  In particular,  for the Schwarzschild
metric solution of the Einstein equations obtained from the usual Einstein-Hilbert action, the  effective action $\Delta S_{\rm g}$
diverges as $(1-2MG/r)^{-2}$ near the Schwarzschild radius $r_S=2MG$, suggesting that including this term and solving the resulting
modified Einstein equations may substantially affect the horizon structure. Thus, a second motivation
of this paper is to study places where the two interpretations of dark energy, either as a ``bare'' cosmological constant arising from Lorentz
invariant vacuum physics, or as
a frame-dependent effect of pre-quantum matter fluctuations, give different results.  The natural place to start such a
study is in the well-defined arena of spherically symmetric metrics, which was initiated in \cite{adlerpaper}.
Our aim in the present paper is to continue the analysis of the spherically symmetric case, both for static and
 time dependent metrics, by deriving and solving
the differential equations governing spherically symmetric
Schwarzschild-like solutions in the preferred rest frame for the effective action.
\subsection{Outline and brief summary}
We proceed now to give an outline of the paper, and a brief summary of the principal results.
 In Sec. II we analyze the equations in standard
 polar coordinates, marshalling both analytic and numerical results.  In Sec. III we study the related
 equations that describe the Schwarzschild-like solutions in isotropic coordinates.  In Sec. IV we extend our
 results to the time-dependent case, focussing on the McVittie Ansatz which, as in the static case, can be
 analyzed by solving ordinary differential equations in a radius-like variable.  In Sec. V we discuss the
 relevance of our results to outstanding issues in black hole physics.  Notational conventions and
 extensions of our discussion are given in Appendices.

 Our results show that, as anticipated, the effective action of Eq. \eqref{eq:action2} substantially changes the horizon structure.
 At macroscopic distances $>> 10^{-17} {\cal M}$ cm from the nominal horizon (with ${\cal M}$ the black hole
 mass in solar mass units) the
 solutions closely approximate the standard Schwarzschild form until cosmological distances are reached. Hence we do not expect significant
 consequences for the  astrophysics of stellar and galactic structure.  Within $10^{-17}{\cal M}$ cm from the nominal horizon, the behavior
 of $g_{00}$ changes, with $g_{00}$ remaining non-vanishing until the internal singularity is reached; this could have consequences, still to be
 explored, for discussions of the black hole ``information paradox''.  We can conclude little from the internal singularity
 itself, since here the approximation of neglecting metric derivatives in the effective action of Eq. \eqref{eq:action1} breaks down.
 At cosmological distances,
 the static solutions have a physical (not a coordinate) singularity, which may be a consequence of using
 the static metric assumption in a cosmological regime where it does not apply. Introducing time dependence significantly
 alters the form of the spherically symmetric solutions.
 Time-dependent solutions  in the McVittie Ansatz are very sensitive to initial values and integration details, but
 subject to uncertainties which we discuss, there may be solutions which have similar horizon behavior to the static case, but in which the large distance physical singularity is pushed out to super-cosmological distances for black hole masses of astrophysical interest.

 \section{Schwarzschild-like solutions in standard polar coordinates}

\subsection{Modified Einstein equations for the static, spherically symmetric metric}

The standard form for the static, spherically symmetric line element is
\begin{equation}\label{eq:polar}
(ds)^2=B(r) (dt)^2 - A(r) (dr)^2 - r^2\big( (d\theta)^2 + \sin^2 \theta (d\phi)^2\big)~~~,
\end{equation}
corresponding to the metric components
\begin{equation}
g_{tt}\equiv g_{00}=B(r)~,~~g_{rr}=-A(r)~,~~g_{\theta \theta}= -r^2~,~~g_{\phi \phi}= -r^2 \sin^2\theta~~~.
\end{equation}
Since we have $g_{0i}=g_{i0}=0\,,~D^i=0$,  we can  use the simplified form of the
induced action given in Eqs. \eqref{eq:action1}--\eqref{eq:action2}.
Substituting $g_{00}=B(r)$, we get
\begin{equation}\label{eq:spher}
\Delta S_{\rm g}=-\frac{\Lambda}{8 \pi G} \int d^4x (^{(4)}g)^{1/2} B(r)^{-2}~~~.
\end{equation}
 Varying the spatial components $g_{ij}$ of the metric, while taking $\delta g_{00}=\delta g_{0i}=0$,
 and writing
 \begin{equation}\label{eq:tdef}
\delta \Delta S_{\rm g}=-\frac{1}{2} \int d^4 x (^{(4)}g)^{1/2} \Delta T^{ij } \delta g_{ij}~~~.
\end{equation}
we find from Eq. \eqref{eq:spher}  that the spatial components  $\Delta T^{ij}$
are given by
\begin{equation}\label{eq:tijeqs}
\Delta T^{ij}=\frac{\Lambda}{8 \pi G} g^{ij}/B(r)^2~,~~~ \Delta T_{ij} = \frac{\Lambda}{8 \pi G}g_{ij}/B(r)^2~~~.
\end{equation}
Adding this to the variation of the Einstein action $S_g$ of Eqs. \eqref{eq:einsteinaction} and \eqref{eq:totalaction},
and equating the total variation to zero, we get the modified
 Einstein equations for $G_{rr}$ and $G_{\theta \theta}$,
\begin{align}\label{eq:withbsq}
&G_{rr}- \frac {\Lambda A(r)}{ B(r)^2}=0~~~,\cr
&G_{\theta \theta} -\frac {\Lambda r^2} { B(r)^2}=0~~~,\cr
\end{align}
with the equation for $G_{\phi \phi}$ proportional to
that for $G_{\theta \theta}$.

From the expressions for $G_{tt}$, $G_{rr}$, and $G_{\theta \theta}$, with $^{\prime}$ denoting $d/dr$,
and with $A\equiv A(r)$ and $B \equiv B(r)$ in Eqs. \eqref{eq:spher1} -- \eqref{eq:bianchi},
\begin{align}\label{eq:spher1}
G_{tt}=&\frac{B}{rA} \left[ -\frac{A^{\prime}}{A}+ \frac{1}{r} (1-A)\right]~~~,\cr
G_{rr}=&-\frac{B^{\prime}}{rB} + \frac{1}{r^2} (A-1)~~~,\cr
G_{\theta \theta}= -&\frac{r^2}{2A}\left[\frac{B^{\prime \prime}}{B} - \frac{B^{\prime}}{2B}
\left(\frac{A^{\prime}}{A}+\frac{B^{\prime}}{B}\right)+\frac{1}{r}\left(-\frac{A^{\prime}}{A}+\frac{B^{\prime}}{B} \right) \right]~~~,\cr
\end{align}
we find the linear relation (the Bianchi identity)
\begin{equation}\label{eq:bianchi}
G_{rr}^{\prime} -\frac{2 A}{r^3} G_{\theta \theta} +\left( \frac{B^{\prime}}{2B}+\frac{2}{r}-\frac{A^{\prime}}{A} \right) G_{rr}
+\frac{A B^{\prime}}{2B^2}G_{tt}=0~~~.
\end{equation}
When $G_{\mu \nu}$ is replaced in this equation by the covariantly conserved $\Delta T_{\mu \nu}$ it must also be satisfied,  so for the
conserving extension $\Delta T_{tt}$ of $\Delta T_{rr}$ and $\Delta T_{\theta \theta}$ defined in \cite{adlerpaper}  we find (see also Appendix B)
\begin{equation}\label{eq:ttt}
\Delta T_{tt}=-\frac {3 \Lambda}{8 \pi G  B}~~~,
\end{equation}
giving as the modified equation for $G_{tt}$
\begin{equation}\label{eq:gtteq}
G_{tt}-\frac {3 \Lambda}{B}=0~~~.
\end{equation}
The differential equations derived in the subsequent analysis proceed from the equations for
$G_{rr}$ and $G_{\theta \theta}$, but we will use the $G_{tt}$ equation in deriving the value
of the curvature scalar $R$.

Proceeding in an analogous way, one can derive the modified equations for a general axially symmetric metric.  These are
given (but not solved) in Appendix D of the arXiv version of this paper (arXiv:1308.1448v3).

\subsection{Comparison with the Schwarzschild-de-Sitter equations and metric}

Before proceeding to study  Eqs. \eqref{eq:withbsq}, we first
examine the analogous equations that would be obtained from an ordinary cosmological
constant \big(that is, Eq. \eqref{eq:lambda0} with $\Lambda_0$ interpreted as the observed cosmological
constant $\Lambda$\big) , which correspond to replacing $B(r)^2$ by unity in Eqs. \eqref{eq:spher} --
\eqref{eq:withbsq}.  This gives the Einstein equations
\begin{align}\label{eq:withoutbsq}
&G_{rr}- \Lambda A(r)=0~~~,\cr
&G_{\theta \theta} -\Lambda r^2=0~~~.\cr
\end{align}
As is well known, this system of equations can be solved by making the
Ansatz $A(r)=C/B(r)$, with $C$ a constant, which implies that $A^{\prime}/A+B^{\prime}/B=0$.
Substituting this, the $G_{\theta \theta}$ equation reduces to
\begin{equation}\label{eq:gth}
B^{\prime \prime}+\frac{2}{r}B^{\prime} +2 \Lambda=0~~~,
\end{equation}
and the $G_{rr}$ equation takes the form
\begin{equation}\label{eq:grr}
B^{\prime} + \frac{1}{r}B + r \Lambda-\frac{C}{r}=0~~~.
\end{equation}
Differentiating Eq. \eqref{eq:grr} once we get
\begin{equation}
B^{\prime\prime}+\frac{1}{r}B^{\prime}-\frac{1}{r^2}B+\Lambda +\frac{C}{r^2}=0~~~,
\end{equation}
which when subtracted from Eq. \eqref{eq:gth} just gives $1/r$ times Eq. \eqref{eq:grr}.
Hence the Ansatz is self-consistent, and it suffices to solve the first order linear equation Eq. \eqref{eq:grr}.  Making the conventional choice $C=1$, and setting the gravitational
constant $G$ to unity so that the
Schwarzschild radius becomes $2M$, this has its solution the Schwarzschild-de-Sitter
metric,
\begin{align}\label{eq:scdesit}
B(r)=&1-\frac{2M}{r}-\frac{1}{3}\Lambda r^2~~~,\cr
A(r)=&1/B(r)~~~.\cr
\end{align}
When $M\Lambda^{1/2}<<1$,  the Schwarzschild horizon for this metric is at $r \simeq 2M$.  There is also a de Sitter horizon at
$r \simeq (3/\Lambda)^{1/2}$, which is a coordinate singularity (but not a physical singularity, since curvature invariants
remain finite there).  For a detailed discussion of the Schwarzschild-de-Sitter metric, see  Gibbons and Hawking \cite{gibbons},
and for an extensive list of references, organized by category, see Podolsk\'y \cite{pod}.

\subsection{Reducing the effective action equations to a second order differential equation for $B(r)$ }

We now return to Eqs. \eqref{eq:withbsq}, for which we cannot use the simplifying Ansatz
$A(r) \propto 1/B(r)$.  We proceed by first solving the $G_{rr}$ equation for $A$,
\begin{equation}\label{eq:solveA}
A=\frac{B^2+rBB^{\prime}}{B^2-\Lambda r^2}~~~,
\end{equation}
from which we find
\begin{equation}\label{eq:solveAprim}
\frac{A^{\prime}}{A}=
\frac{3BB^{\prime}+r[(B^{\prime})^2+B B^{\prime \prime}]}{B^2+rBB^{\prime}}
-\frac{2(BB^{\prime}-\Lambda r)}{B^2-\Lambda r^2}~~~.
\end{equation}
Substituting these equations into the $G_{\theta \theta}$ equation, and doing some
algebraic simplification, we get the second order, nonlinear differential ``master equation'' for $B$,
\begin{equation}\label{eq:masterb1}
B^{\prime\prime}+\frac{2}{r} B^{\prime} +\frac{2 \Lambda (B^{\prime}r+B)(B^{\prime}r+3B)}
{B(B^2-\Lambda r^2)}=0~~~.
\end{equation}
We will employ various different forms of this exact master equation in the ensuing discussion.

Often in the literature, the $A$ and $B$ functions for the static spherically symmetric metric are rewritten in exponential form as
\begin{equation}
B\equiv \exp(2F)~~~,~ A\equiv \exp(2H)~~~,
\end{equation}
where $F$ and $H$ are not restricted to be real-valued (allowing $A$ and $B$ to be negative when $2F$ and $2G$ develop logarithmic branch cuts, as within the horizon of the standard Schwarzschild solution).
In terms of the functions $F,\,H$, the equation relating $A$ to $B$ is
\begin{equation}\label{eq:HfromF}
\exp(2H)=\frac{\exp(4F) (1+2r F^{\prime})}{\exp(4F)-\Lambda r^2}~~~,
\end{equation}
and the differential equation for $F$ is
\begin{equation}\label{eq:masterf}
F^{\prime \prime} +2 (F^{\prime})^2+ \frac{2}{r} F^{\prime}
+\frac{\Lambda (2 r F^{\prime}+1)(2rF^{\prime}+3)}{\exp(4F)-\Lambda r^2}=0~~~.
\end{equation}

\subsection{Exact particular solutions of the master equation}

We have not succeeded in finding an analytic general solution to the master equation,
but it is easy to find two exact particular solutions.  The first is obtained by noting
that $B^{\prime\prime}+(2/r) B^{\prime}$ and $B^{\prime}r+B$ both vanish for $B=K/r$ with any value of the constant of integration $K$.  Hence
\begin{equation}
B= K/r
\end{equation}
is an exact particular solution of the master equation, and corresponds to the leading term of the small $r$ behavior exhibited below
in Eq. \eqref{eq:leadingsmallx}.
The second is obtained by
trying the Ansatz $B=C \Lambda^{1/2} r$.  The $\Lambda$ and $r$ dependence then cancel
out of the master equation, leaving an algebraic equation
\begin{equation}
2 C + \frac{2(2C)(4C)}{C (C^2-1)}=0~~~,
\end{equation}
which simplifies to $C^2+7=0$.
Thus
\begin{equation}\label{eq:complexsoln}
B=\pm (7\Lambda)^{1/2}  i r
\end{equation}
is a purely imaginary-valued particular solution of the master equation.  Equation \eqref{eq:complexsoln} does not play a role in
our subsequent analysis, which focuses on physically relevant solutions that are real-valued outside a finite radius.  To study such
physical general solutions, we resort in the following sections to analytic approximation and numerical methods.

\subsection{Leading perturbative correction to the Schwarzschild metric}

Since the observed cosmological constant $\Lambda$ is very small, it is useful to develop
Eqs. \eqref{eq:masterb1} and \eqref{eq:masterf} in a perturbation expansion in $\Lambda$.
To do this we write (again setting the gravitational constant $G$ to unity)
\begin{align}\label{eq:expand}
B=&B^{(0)}+ \Lambda B^{(1)}+...,\cr
F=&F^{(0)}+ \Lambda F^{(1)}+....,\cr
B^{(0)}=&1-2M/r~~~,~F^{(0)}=\frac{1}{2}\log(1-2M/r)~~~,\cr
B^{(1)}=&2(1-2M/r)F^{(1)}~~~,\cr
\end{align}
with $B^{(1)}$ and $F^{(1)}$ coefficients of the terms of first order in $\Lambda$.
Substituting these expansions into Eqs. \eqref{eq:masterb1} and \eqref{eq:masterf}
we find the following equations governing the first order perturbations.
For $B^{(1)}$ we have
\begin{equation}
r B^{(1)\prime\prime} + 2 B^{(1)\prime} +\frac{2 r^3 (3r-4M)}{(r-2M)^3}=0~~~,
\end{equation}
and for $F^{(1)}$ we have
\begin{equation}
(1-2M/r)F^{(1)\prime\prime}+(2/r)F^{(1)\prime} + \frac{ (3-4M/r)}{(1-2M/r)^3}=0~~~.
\end{equation}
These equations are readily integrated; the $F^{(1)}$ equation is easier to integrate
``by hand'', but we have also solved both using Mathematica and cross-checked the results.
For $F^{(1)}$ we find
\begin{align}\label{eq:ffinal}
F^{(1)}=&- \left\{ \frac{1}{2}(r-2M)^2  +\frac{8M^4}{(r-2M)^2}+10 M (r-2M) -\frac{48M^3}{r-2M}\right.\cr
&\left.+48M^2{\log}[(r-2M)/M_1]-\frac{48M^3{\log}[(r-2M)/M_2]}{r-2M} \right\} ~~~,\cr
\end{align}
 where we have introduced the constants of integration as scale factors $M_1, \, M_2$ in the logarithms. (These
 correspond to small renormalizations of the 1 and $2M/r$ terms in the zeroth order solution, and could be omitted.)
 The corresponding expression for $B^{(1)}$ follows from the final line of Eq. \eqref{eq:expand}.  Writing
\begin{align}
H=&H^{(0)}+\Lambda H^{(1)}+....,\cr
H^{(0)}=&-\frac{1}{2}\log(1-2M/r)~~~,\cr
\end{align}
expanding Eq. \eqref{eq:HfromF} to get
\begin{equation}
H^{(1)}=(r-2M)F^{(1)\prime}+\frac{r^4}{2}\frac{1}{(r-2M)^2}~~~,
\end{equation}
and using Eq. \eqref{eq:ffinal}, we get the corresponding expression for
$H^{(1)}$,
\begin{equation}
H^{(1)}=-\left\{(r-2M)^2-\frac{(1/2)r^4+16M^4}{(r-2M)^2}+10M(r-2M)
+48M^2+\frac{48M^3 \log[(r-2M)/M_2]}{r-2M}\right\}~~~.
\end{equation}

Near the Schwarzschild metric horizon at $r=2M$, the leading corrections to $A$ and $B$ are
\begin{align}\label{eq:horizon}
B =& (1-2M/r)\left( 1-\Lambda \frac{16M^4}{(r-2M)^2}+O(\Lambda^2) \right)~~~,\cr
A =& (1-2M/r)^{-1} \left( 1+\Lambda \frac{48M^4}{(r-2M)^2}+O(\Lambda^2) \right)~~~,\cr
\end{align}
while for large $r$, the corresponding leading corrections are
\begin{align}\label{eq:larger}
B =& (1-2M/r)\left(1-\Lambda r^2+ O(\Lambda^2)\right)~~~,\cr
A =& (1-2M/r)^{-1} \left(1-\Lambda r^2 +O(\Lambda^2)\right)~~~.\cr
\end{align}

\subsection{Dimensionless forms of the master equation}

To further analyze the master equation, it is convenient to put it in dimensionless
form by scaling out  the dimensional constant $\Lambda$. Defining
\begin{align}\label{eq:dimensionless}
x=&\Lambda^{1/2}r~~~,\cr
B(r)=& B(x/\Lambda^{1/2})\equiv b(x)~~~,
\end{align}
we find that $b(x)$ satisfies the differential equation, with $^{\prime}$  denoting now $d/dx$,
\begin{equation}\label{eq:masterb}
b^{\prime\prime} + \frac{2}{x}b^{\prime}+\frac{2(xb^{\prime}+b)(xb^{\prime}+3b)}{b(b^2-x^2)}=0~~~.
\end{equation}
This equation can be rewritten in the form
\begin{align}\label{eq:pole}
b^{\prime \prime}=&L(b,x)(b^{\prime})^2+M(b,x)b^{\prime} + N(b,x)~~~,\cr
L(b,x)=& \frac{2}{b}-\frac{1}{b-x}-\frac{1}{b+x}~~~,\cr
M(b,x)=& -\frac{2}{x}+4\left(\frac{1}{b+x}-\frac{1}{b-x}\right)~~~,\cr
N(b,x)=&-3\left(\frac{1}{b-x}+\frac{1}{b+x}\right)~~~,\cr
\end{align}
which exhibits the pole structure of the coefficients.  An examination of the numerators
in Eq. \eqref{eq:pole} shows that they do not have Painlev\'e equation form \cite{ince};
this will be confirmed below when we show that the solution has a movable square root
branch point.

Alternative forms of the master equation can be obtained by various changes of
dependent variable.  If we write $w=xb^{\prime}+b$, then the equation takes the form
\begin{equation}
w^{\prime}+\frac{2xw(w+2b)}{b(b^2-x^2)}=0~~~,
\end{equation}
which expresses the master equation as a system of two first order equations.
If we write $b(x)=xf(x)$, the equation takes the form
\begin{equation}
x^2f^{\prime\prime}+4xf^{\prime}+2f+\frac{2(xf^{\prime}+2f)(xf^{\prime}+4f)}{f(f^2-1)}=0~~~,
\end{equation}
while writing $c(x)=xb(x)$ the equation becomes
\begin{equation}\label{eq:cequ}
c^{\prime \prime}+ \frac{2 x^3 c^{\prime}(xc^{\prime}+2c)}{c(c^2-x^4)}=0~~~.
\end{equation}

Returning to Eq. \eqref{eq:masterb}, the master equation simplifies in two limiting
regimes.  In the large $x$ regime, where $x^2>>b^2$, the equation simplifies to
\begin{equation}\label{eq:bmasterlarge}
b^{\prime\prime} + \frac{2}{x}b^{\prime}-\frac{2(xb^{\prime}+b)(xb^{\prime}+3b)}{bx^2}=0~~~,
\end{equation}
which can be rewritten as
\begin{equation}\label{eq:large1}
b^{\prime\prime}=\frac{2}{b}(b^{\prime})^2+ \frac{6}{x}b^{\prime} +\frac{6}{x^2}b~~~.
\end{equation}
In the small $x$ regime, where $x^2<<b^2$, it simplifies to
\begin{equation}\label{eq:bmastersmall}
b^{\prime\prime} + \frac{2}{x}b^{\prime}+\frac{2(xb^{\prime}+b)(xb^{\prime}+3b)}{b^3}=0~~~,
\end{equation}
which can be rewritten as
\begin{equation}\label{eq:small1}
b^{\prime\prime}=-\frac{2x^2}{b^3}(b^{\prime})^2 -2\left(\frac{1}{x}+\frac{4x}{b^2}\right)
b^{\prime}-\frac{6}{b}~~~.
\end{equation}

In between the large $x$ and small $x$ regimes, there is a regime where $b \simeq x$.
To study the behavior here, we write $b=x+d(x)$, and substitute into Eq. \eqref{eq:masterb},
giving
\begin{equation}\label{eq:ford}
d^{\prime \prime}+ \frac{2}{x}(1+d^{\prime})+\frac{2(2x+xd^{\prime}+d)(4x+xd^{\prime}+3d)}
{d(x+d)(2x+d)}=0~~~.
\end{equation}
\subsection{Large and small $x$ behavior}

We proceed now to examine the large and small $x$ behavior.  By making the change of
dependent variable \cite{ince} $b=1/W$, Eq. \eqref{eq:large1} is transformed to the linear equation
\begin{equation}
W^{\prime\prime}=\frac{6}{x}W^{\prime}-\frac{6}{x^2}W~~~,
\end{equation}
which has the general solution
\begin{equation}
W=C_1x +C_2 x^6~~~,
\end{equation}
corresponding to
\begin{equation}\label{eq:largexact}
b=(C_1x+C_2x^6)^{-1}  ~~~~~~{\rm large~x}~~~.
\end{equation}
Since this shows that $b \to 0$ as $x \to \infty$, the condition $x^2>>b^2$ used to
simplify the master equation in the large $x$ regime is obeyed.

We have not been able to find an exact integral for Eq. \eqref{eq:small1} (or for the
general master equation), so we look for the leading power behavior.  Substituting
the Ansatz $b=Cx^{\lambda}$ into Eq.  \eqref{eq:bmastersmall}, we find
\begin{equation}
(\lambda+1)[C\lambda x^{\lambda-2}+2C^{-1}(\lambda+3)x^{-\lambda}]=0~~~.
\end{equation}
Hence $\lambda =-1$ gives one possible small $x$ behavior.  If $\lambda-2 < -\lambda$,
that is, $\lambda<1$.
the first term in brackets dominates at $x=0$, and gives the condition $\lambda=0$,
which is consistent with the condition $\lambda <1$, giving a second possible small $x$
behavior.
If $\lambda>1$, the second term in brackets dominates at $x=0$, and gives the
condition $\lambda=-3$, which is not consistent with $\lambda >1$, and so does not
give an additional small $x$ behavior, which is expected since a second order equation
should have only two undetermined constants of integration.  So the leading small
$x$ behavior is
\begin{equation}\label{eq:leadingsmallx}
b=C_3/x +C_4~~~~~~{\rm small~x}~~~,
\end{equation}
which is consistent with the condition $b^2>>x^2$ used to simplify the master equation
in the small $x$ regime.

\subsection{Behavior of $b$  in $0<x<\infty$}

We next examine the behavior of the solution for finite, nonzero values of $x$.
Starting from the general form of the master equation given in Eq. \eqref{eq:masterb}, we shall demonstrate several things:  (i) $b$ cannot vanish in the interval $0<x<\infty$,
(ii) $b$ cannot become infinite in the interval $0<x<\infty$, (iv) $b$ develops a square
root branch point where $b \simeq x$, and
(iv) $b$ can have minima or maxima in the interval $0<x<\infty$, and must have
negative slope at points of inflection.

To study the possible presence of zeros of $b$, we note that in the vicinity
of a zero of $b$ at $x=a$ we must have $x^2>>b^2$, and so we can use the large $x$
approximation to the master equation given in Eq. \eqref{eq:large1}.  Substituting
the Ansatz $b=C(x-a)^{\lambda}$, with $\lambda>0$, we find
\begin{equation}
\lambda(\lambda-1)(x-a)^{\lambda-2}=2\lambda^2(x-a)^{\lambda-2} +(6/a)\lambda (x-a)^{\lambda-1} +(6/a^2)(x-a)^{\lambda}~~~,
\end{equation}
in other words
\begin{equation}
-\lambda(\lambda+1) (x-a)^{\lambda-2}= (6/a) \lambda (x-a)^{\lambda-1} +(6/a^2)(x-a)^{\lambda}~~~.
\end{equation}
Vanishing of the leading term requires $\lambda(\lambda+1)=0$, so that either $\lambda=0$ or $\lambda=-1$, both of which are inconsistent with
the condition $\lambda>0$ needed for a zero of $b$ at $x=a$.  Hence $b$ cannot vanish
in the interval $0<x<\infty$.   This conclusion also follows from the exact solution to Eq. \eqref{eq:large1}
given in Eq. \eqref{eq:largexact}, which cannot develop a zero for finite values of $x$.

To study the possible presence of infinities of $b$, we note that in the vicinity of
an infinity of $b$ at $x=a$, we must have $b^2>>x^2$, and so we can use the small
$x$ approximation to the master equation given in Eq. \eqref{eq:small1}.  Substituting
the Ansatz $b=C(x-a)^{\lambda}$, with $\lambda<0$, we find
\begin{align}
&C[\lambda(\lambda-1)(x-a)^{\lambda-2} +(2/a) (x-a)^{\lambda-1}]\cr
&+\frac{2}{C(x-a)^{3\lambda}}[a\lambda(x-a)^{\lambda-1}+(x-a)^{\lambda}]
[a\lambda(x-a)^{\lambda-1}+3(x-a)^{\lambda}]=0~~~,\cr
\end{align}
in other words
\begin{align}
&C[\lambda(\lambda-1)(x-a)^{\lambda-2} +(2/a) (x-a)^{\lambda-1}]\cr
&+\frac{2}{C}[a^2 \lambda^2 (x-a)^{-\lambda-2}+4a\lambda(x-a)^{-\lambda-1}
+3(x-a)^{-\lambda}]~~~.\cr
\end{align}
Since by assumption $\lambda<0$, the dominant term is the one with exponent $\lambda-2$,
giving the condition $\lambda (\lambda-1)=0$.  This requires either $\lambda =0$ or
$\lambda=1$, both of which are inconsistent with the condition $\lambda<0$ needed for
an infinity of $b$ at $x-a$.  Hence $b$ cannot become infinite in the interval
$0<x<\infty$.

Next, we study the behavior of $b$ near where $b(x)$ crosses $x$, using Eq. \eqref{eq:ford}.
Substituting the Ansatz $d=C(x-a)^{\lambda}$, with $\lambda \geq 0$ (since $b$ cannot be
infinite at $x=a$), we find
\begin{equation}\label{eq:leading}
C[\lambda (\lambda-1)+ \lambda^2] (x-a)^{\lambda-2}+ O\big((x-a)^{\lambda-1}\big)
+O\big((x-a)^{\lambda}\big) + O\big((x-a)^{-1}\big)+O\big((x-a)^{-\lambda}\big)+O(1)=0~~~.
\end{equation}
Equating the coefficient of the leading term $(x-a)^{\lambda-2}$ term to 0, we find
the condition $\lambda (2\lambda-1)=0$.  The root $\lambda=0$ corresponds to $b(x)=x+C$,
which does not cross $x$, while the root $\lambda=1/2$ corresponds to a crossing behavior
of
\begin{equation}\label{eq:crossing}
b(x)=x+C(x-a)^{1/2}~~~,
\end{equation}
which has a square root branch cut starting at $x=a$. So for $x<a$, the solution $b$ becomes
complex, with a two-sheeted Riemann sheet structure.  As a check, we note that for $\lambda =1/2$, all of the terms indicated as subdominant in Eq. \eqref{eq:leading} have
exponents greater than $\lambda-2=-3/2$, and so are indeed subdominant.  An alternative
way to get the same results is to make the Ansatz that at $x=a$, d is finite but $d^{\prime}$
and $d^{\prime \prime}$ are infinite.  Then Eq. \eqref{eq:ford} simplifies to the
differential equation
\begin{equation}
d^{\prime \prime}+ \frac{ (d^{\prime})^2}{d}=0~~~,
\end{equation}
that is $(dd^{\prime})^{\prime}=0$, which has the general solution
$d=(C_2+C_1x)^{1/2}$, again showing that there is a square root branch cut starting
where $b(x)$ crosses $x$.

We note, however, from Eq. \eqref{eq:masterb}, that if $b(x)$ crosses $x$ at a point where the slope $b^{\prime}(x)=-1$,
then the vanishing numerator factor $xb^{\prime}+b$ cancels the vanishing denominator factor $b(x)-x$, and the solution
can be regular.  This behavior is in fact realized in the large $x$ portion of the solution, where $b(x)$ is decreasing
from its peak value.

Next, we show that $b$ can have local minima or maxima in the interval $0<x<\infty$. At
a generic point $x=a$ where $b$ has zero slope, the power series expansion for $b$ takes the form
\begin{equation}\label{eq:expansion}
b=c+d(x-a)^2+e(x-a)^3+f(x-a)^4+g(x-a)^5+...~~~.
\end{equation}
Multiplying Eq. \eqref{eq:masterb} through by $xb(b^2-x^2)$, which turns it into a polynomial equation, substituting Eq. \eqref{eq:expansion} for $b$ and using Mathematica to expand
in powers of $x-a$, we get the conditions
\begin{align}
d=&3c/(a^2-c^2)~~~,\cr
e=&[c+d(a^2+c^2)]/[a(a^2-c^2)]~~~,\cr
f=&[a^3d^2+5acd+ac^2d^2+2c^2e]/[2ac(a^2-c^2)]~~~,\cr
&{...........}~~~.\cr
\end{align}
So $a$ and $c$ function as the two expected constants of integration, and the rest
of the coefficients are determined by them.  Note that if $c=0$, all subsequent coefficients
vanish, in agreement with the fact that $b$ cannot vanish at a finite point $x=a$. This
conclusion still follows if a linear term $\ell (x-a)$ is added to the series expansion:
if $a\neq 0$, vanishing of $c$ implies vanishing of the power series solution.  Note, however, that if $a=0$, a zero crossing is allowed, as shown explicitly by the exact, complex-valued
solution of Eq. \eqref{eq:complexsoln}.

Finally, we use the same series expansions to show that at points of inflection where
$b^{\prime \prime}=0$, the slope $b^{\prime}<0$.  To see this, substitute $b=c+{\ell}(x-a)a$
into Eq. \eqref{eq:masterb}, giving at $x=a$ the quadratic equation for $\ell$
\begin{equation}
\ell^2+[(c/a)^{3}+3(c/a)]\ell+3(c/a)^2=0~~~.
\end{equation}
Since all coefficients in this equation are positive, it cannot have roots with positive or
zero $\ell$, so the roots must have $\ell$ negative (or complex).  This is a clue to what
we found above, that where the small $x$ regime merges into the solution for larger $x$,
$b^{\prime\prime}$ must have a singularity, since a point of inflection with $b^{\prime \prime}=0$ and $b^{\prime}>0$ is not allowed.

\subsection{Qualitative form of $B$ and $b$, and discussion}

Drawing on the results of the previous subsections, we can give a qualitative sketch
of the form of the solution to the master equation.  We assume that the two
constants of integration are chosen so that when $\Lambda \to 0$,
the solution approaches the Schwarzschild solution $B=C-2M/r$ with the conventional choice
of the first constant of integration $C=1$, leaving only the second constant of integration $2M$ as a parameter.  We first give our sketch in terms of $g_{00}(r)=B$ and $r$, and then restate it  in terms of $g_{00}(x)=b$ and the rescaled radius variable $x$.  Because
$B$ can have no zeros at finite radius, $B$ is always positive, and we assume that are no
``extra wiggles'' resulting in more maxima/minima than required by our above analysis.  In particular, we assume (as is verified
by the numerical solution) that when $b(x)$ crosses $x$ at large $x$, the slope is $b^{\prime}(x)=-1$, allowing the soution
to be regular.

There are four distinct regimes, moving inward from $r=\infty$:

(I) The region of large $r$, where $\Lambda r^2 = O(1)$, that is, $r \sim \Lambda^{-1/2}$.
Moving out from the Schwarzschild-like region (II), $B$ starts to fall off as $1-\Lambda r^2$,
faster than the Schwarzschild-de-Sitter behavior $1-\Lambda r^3/3$.  However, the vanishing of $B$
is pushed out to $r=\infty$, and asymptotically $B$ falls off as a  positive constant times $r^{-6}$.

(II) The Schwarzschild-like, or Newtonian regime $\Lambda^{-1/2}>>r>>2M$.  Here
$B \simeq 1-2M/r$, with  perturbative corrections as derived in Sec. IIE.  Going from
region (II), where $B$ is increasing with increasing radius, to region (I), where $B$ decreases
with increasing radius, there must be a local maximum of $B$.

(III) The immediate vicinity of the Schwarzschild horizon, which according to the most
singular term of the perturbative expansion is  $\Lambda M^4/(r-2M)^2 = O(1)$, that is $|r-2M| = O(\Lambda^{1/2} M^2)$ ($\sim 10^{-17}\,{\rm cm}$ for a solar mass black hole, and $\sim 10^{-7}\,{\rm cm}$ for a $10^{10}$ solar mass black hole). Here $B$ becomes complex, with a square root branch point,
and $|B|$ is repelled from the $B=0$ axis, since $B$ cannot be zero, and turns upward, so in
region (III) or in the transition to region (IV), there must be a local minimum of $|B|$.

(IV)  The regime of small $r$, where $r<<2M$.   Here $B$ is complex and $|B|$ approaches $+\infty$ as a constant times $r^{-1}$ as $r \to 0$.

In terms of the dimensionless radius variable $x$, the four regimes are:

(I) The large $x$ regime $x=O(1)$.

(II) The Schwarzschild-like regime $1>>x>>\Lambda^{1/2} M$.

(III) The horizon regime $|x-2\Lambda^{1/2} M| =O(\Lambda M^2)$.

(IV) The small $x$ regime, $x<<\Lambda^{1/2} M$.

These are sketched qualitatively in Fig. 1

\begin{figure}[t]
\centering
\includegraphics{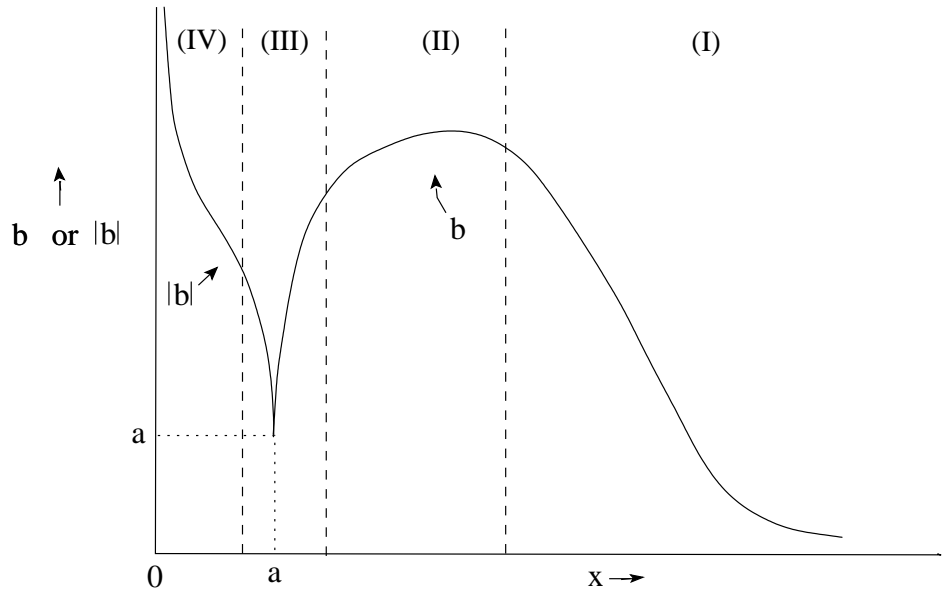}
\caption{Qualitative sketch (not to scale) of $b$ and $|b|$ versus $x$ \big(these are the dimensionless polar coordinate variables of Eq. \eqref{eq:dimensionless}\big).}
\end{figure}

\begin{figure}[b]
\centering
\includegraphics{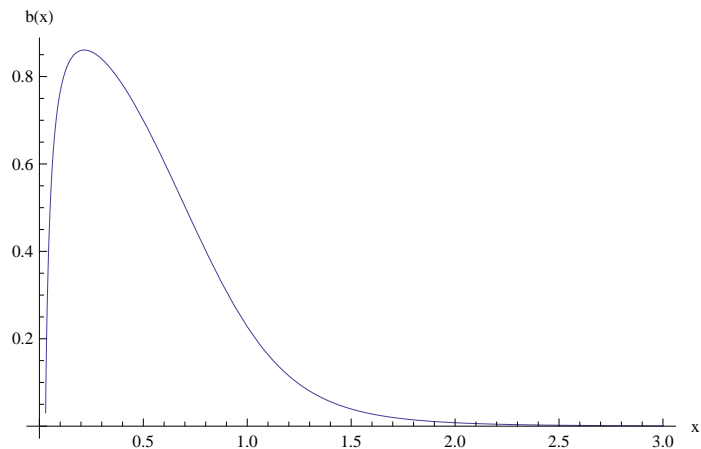}
\caption{Numerical result for $b$ versus $x$ \big(these are the dimensionless polar coordinate variables of Eq. \eqref{eq:dimensionless}\big) with $M\Lambda^{1/2}=10^{-2}$,
on a linear plot.}
\end{figure}

\subsection{Numerical results for $b$}

Numerical results for $b$ obtained for $x>a$ by solving the
master equation using Mathematica are given in Fig. 2 (linear plot) and Fig. 3 (log log plot),
for a mass parameter $M\Lambda^{1/2}=10^{-2}$.  The mass was input through initial conditions
imposed in region II,
\begin{align}\label{eq:initcond}
x_0=&(M \Lambda^{1/2})^{1/3}~~~,\cr
b(x_0)=&1-2 M \Lambda^{1/2}/x_0-x_0^2~~~,\cr
b'(x_0)=&0~~~,\cr
\end{align}
with $x_0$ corresponding to the point at which the the rescaled large $r$ solution of
Eq. \eqref{eq:larger},
\begin{equation}\label{eq:matchpoint}
b(x) =1-2M\Lambda^{1/2}/x-x^2
\end{equation}
is a maximum.  We found the structure of the solution to be insensitive to the
point at which the initial conditions on $b$ and $b'$ were applied.  One can
clearly see the cusp at the point $x=a\simeq 0.0301315$, where the solution is
approximated by
\begin{equation}\label{eq:cuspapprox}
b(x)\simeq a+1.2761 (x-a)^{1/2}~~~.
\end{equation}
Where the descending solution $b$ crosses the line $b=x$, the
slope $b'(x)=-1$, which is what allows the solution to be regular.
The log log plot clearly shows
the $x^{-6}$ behavior at large $x$. As the rescaled mass
$M\Lambda^{1/2}$ approaches zero, the solution approaches a universal mass-independent curve,
given in Fig. 4, which has large x behavior $b(x)\simeq 0.777 x^{-6}$.

\begin{figure}[t]
\centering
\includegraphics{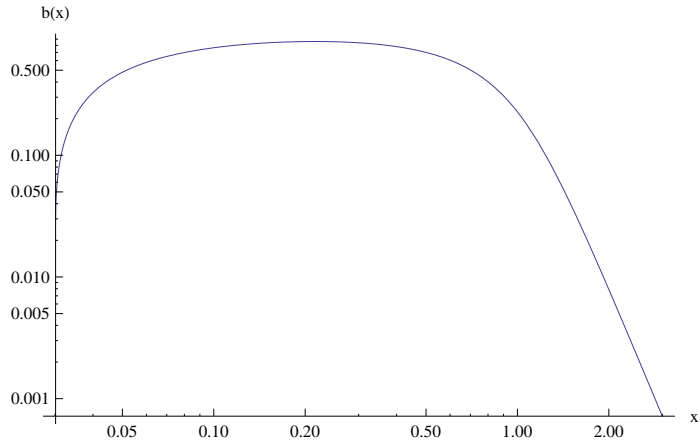}
\caption{Numerical result for $b$ versus $x$ \big(these are the dimensionless polar coordinate variables of Eq. \eqref{eq:dimensionless}\big) with $M\Lambda^{1/2}=10^{-2}$,
on a log log plot.}
\end{figure}

\begin{figure}[t]
\centering
\includegraphics{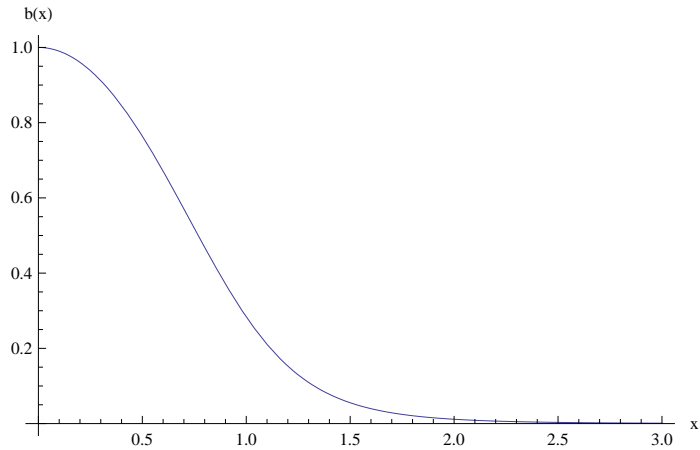}
\caption{Numerical result for $b$ versus $x$ \big(these are the dimensionless polar coordinate variables of Eq. \eqref{eq:dimensionless}\big) with $M\Lambda^{1/2}=0$.}
\end{figure}

\subsection{Vanishing of the curvature scalar $R$}

From the numerical work, we found strong evidence that the curvature scalar $R$ for the master equation solution is
identically zero.  We give here two algebraic proofs of this interesting result.

The first starts from the expression for $-R$ in terms of the Einstein tensor
\begin{equation}\label{eq:req}
-R=g^{\mu\nu}G_{\mu\nu}=g^{tt}G_{tt}+g^{rr}G_{rr}+2g^{\theta\theta}G_{\theta\theta}~~~,
\end{equation}
where we have used the fact that $g^{\phi\phi}G_{\phi\phi}=g^{\theta\theta}G_{\theta\theta}$.
Substituting
\begin{equation}\label{eq:inversemetric}
g^{tt}=1/B(r)~,~~g^{rr}=-1/A(r)~,~~g^{\theta\theta}=-1/r^2~~~,
\end{equation}
together with Eqs. \eqref{eq:withbsq} and \eqref{eq:gtteq}, we get
\begin{equation}\label{eq:req1}
 -R=3\Lambda/B^2 -\Lambda/B^2-2\Lambda/B^2=0~~~.
\end{equation}

A second way to see that $R=0$ is to calculate $R$ in terms of the function $c(x)=xb(x)$
introduced in Eq. \eqref{eq:cequ},
\begin{equation}\label{eq:rinc}
R/\Lambda=\frac{ 3 c(x) -x c^{\prime}(x)} {2 x^2 c(x)^3 c^{\prime}(x)^2}
 [2 x^3 c^{\prime}(x)
  (2 c(x) + x c^{\prime}(x))
  + c(x) (-x^4 + c(x)^2)c^{\prime\prime}(x)]
  ~~~.
\end{equation}
Rewriting the Eq. \eqref{eq:cequ} as
\begin{equation}\label{eq:cequ1}
{c(c^2-x^4)}c^{\prime \prime}+ 2 x^3 c^{\prime}(xc^{\prime}+2c)=0~~~,
\end{equation}
we see that this equation is just a factor of Eq. \eqref{eq:rinc},
and so $R$ vanishes.

\subsection{The cusp is a coordinate singularity, but not a physical singularity}

To study the nature of the cusp at $x=a$, we solve the master equation in a power series
expansion around $x=a$,
\begin{equation}\label{eq:expansion1}
b=a+c(x-a)^{1/2}+d(x-a)+e(x-a)^{3/2}+...~~~.
\end{equation}
The first two coefficients $a$ and $c$ are the two constants of integration, and
the remaining coefficients are determined in terms of them, for example
\begin{equation}\label{eq:dterm}
d=\frac{c^2}{2a}-3~~~.
\end{equation}
From these expansions we have calculated $R_{\mu\nu}R^{\mu\nu}$ and
$R_{\mu\nu\lambda\sigma}R^{\mu\nu\lambda\sigma}$ in expansions around the
cusp,
\begin{align}
R_{\mu\nu}R^{\mu\nu}=&12\frac{\Lambda^2}{a^4}-48  \frac{\Lambda^2}{a^5} c (x-a)^{1/2}+...~~~,\cr
R_{\mu\nu\lambda\sigma}R^{\mu\nu\lambda\sigma}=&72\frac{\Lambda^2}{a^4}-192 \frac{\Lambda^2}{a^5} c  (x-a)^{1/2}+...~~~.\cr
\end{align}
The first of these follows from the Einstein equations by a calculation similar to that used above to show
that $R=0$,
\begin{align}\label{eq:squarecalc}
R_{\mu\nu}R^{\mu\nu}=&G_{\mu\nu}G^{\mu\nu}  \cr
=&(g^{tt}G_{tt})^2+(g^{rr}G_{rr})^2+2(g^{\theta\theta}G_{\theta\theta})^2=12\frac{ \Lambda^2}{B^4}\cr
=&12\frac{\Lambda^2}{a^4}-48  \frac{\Lambda^2}{a^5} c (x-a)^{1/2}+...~~~;
\end{align}
the second requires a lengthier calculation.
We similarly find that the orthonormal frame curvature tensor components have non-vanishing terms
$(x-a)^{1/2}$ in their expansions near the cusp.  Hence these quantities all have coordinate singularities
at the cusp, since $(d/dx) (x-a)^{1/2}=(1/2) (x-a)^{-1/2}$ is infinite at the cusp.

However, letting $D$ denote the proper distance, we have
\begin{align}\label{eq:dDdef}
(dD)^2=&-(ds)^2=A (dr)^2 =\frac{A}{\Lambda} (dx)^2~~~,\cr
d/dD=&\frac{\Lambda^{1/2}}{A^{1/2}} d/dx~~~.\cr
\end{align}
From the series expansion for b, we find from Eq. \eqref{eq:solveA} that
\begin{align}\label{eq:Aformula}
A=&\frac{a}{4(x-a)}[1+\frac{c}{a}(x-a)^{1/2}+O(x-a)]~~~,\cr
A^{1/2}=&\frac{a^{1/2}}{2(x-a)^{1/2}}[1+\frac{c}{2a}(x-a)^{1/2}+O(x-a)]~~~.\cr
\end{align}
Hence
\begin{equation}
d/dD=\frac{2\Lambda^{1/2}}{a^{1/2}}[(x-a)^{1/2}+O(x-a)] d/dx~~~,
\end{equation}
and so $d(R_{\mu\nu}R^{\mu\nu}) /dD$ is finite at $x=a$, and similarly
for other physical curvature quantities.
This argument is easily extended,
by induction, to show that for all $n$,  $(d/dD)^n$ of all physical quantities is finite
at $x=a$,  and so the coordinate singularity at $x=a$ is not a physical
singularity.

\subsection{Physical singularity at $x=\infty$}

From Eq. \eqref{eq:squarecalc}, together with the $x^{-6}$ behavior of $b(x)$ at infinity, we see that
the curvature scalar $R_{\mu\nu}R^{\mu\nu}$ becomes infinite at $x=\infty$,
\begin{equation}
R_{\mu\nu}R^{\mu\nu}= \frac{12 \Lambda^2}{B^4}
\simeq \frac{12 \Lambda^2}{(0.78)^4}x^{24} \simeq 33 \Lambda^2 x^{24}~~~,
\end{equation}
where we have used the asymptotic form of the $M\Lambda^{1/2}\to 0$ limiting solution of Fig. 4.  This physical
singularity is only a {\it finite} proper distance $D$ from the origin, as seen by using
Eq. \eqref{eq:dDdef} which gives
$D=\Lambda^{-1/2}\int dx A(x)^{1/2}$.
Using the limiting solution  to evaluate the integral, we
get
\begin{align}\label{eq:Dcalc}
\int_0^{\infty} dx A(x)^{1/2}=&0.927371~~~,\cr
D=& 0.927371 \Lambda^{-1/2} ~~~.\cr
\end{align}
So recalling that $\Lambda = 3 H_0^2 \Omega_{\Lambda} \simeq 2 H_0^2$, with
$H_0$ the Hubble constant and $\Omega_{\Lambda} \simeq 0.7$ the dark
energy fraction, we see that
\begin{equation}
D\simeq 0.66 H_0^{-1}~~~,
\end{equation}
indicating that the physical singularity lies at a cosmological proper distance
from the origin.

\section{Schwarzschild-like solutions in isotropic coordinates}

We turn next to an examination of static, Schwarzschild-like solutions  in isotropic coordinates.  These coordinates are of intrinsic interest in their own right, and are also needed below in formulating the McVittie Ansatz used
to discuss the time-dependent case. We write the isotropic coordinate line element in the form
\begin{equation}\label{eq:iso}
(ds)^2=\frac{B[r]^2}{A[r]^2} (dt)^2-\frac{A[r]^4}{r^4} [(dr)^2 + r^2 (d\theta)^2 + \sin^2(\theta) (d\phi)^2)] ~~~,
\end{equation}
{\bf where the use of square brackets indicates that these are different functions $A,B$  of a different radial variable from those appearing in Eq. \eqref{eq:polar}.}
The effective action Einstein equations for $G_{rr}$ and $G_{\theta \theta}$ now read
\begin{align}\label{eq:einspatial}
G_{rr}=&2\frac{B  A^{\prime} + A B^{\prime}-2 r A^{\prime} B^{\prime}} {rAB}=-\Lambda \frac{g_{rr}}{g_{00}^2}=\Lambda \frac{A^8}{r^4 B^4}~~~,\cr
G_{\theta\theta}=&-r\frac{B  A^{\prime}+ A B^{\prime}-2 r A^{\prime} B^{\prime}+r(B A^{\prime \prime}+A B^{\prime \prime})}{AB}=-\Lambda \frac{g_{\theta\theta}}{g_{00}^2}=\Lambda \frac{A^8}{r^2 B^4}~~~,\cr
\end{align}
with the equation for $G_{\phi\phi}$ proportional to that for $G_{\theta\theta}$.  Using the Bianchi identity for this metric,
or a simpler covariant conservation argument that applies for static, diagonal metrics given in Appendix B,
we find the Einstein equation for $G_{tt}$ to be
\begin{equation}\label{eq:eintime}
G_{tt}=4r^4\frac{B^2A^{\prime\prime}}{A^7}=\Lambda \frac{3}{g_{00}}=3\Lambda \frac{A^2}{B^2}~~~.
\end{equation}
From the previous two equations we find that the curvature scalar $R$ is
\begin{equation}\label{eq:riso}
R=-2 r^4 \frac{3 B A^{\prime\prime}  + A B^{\prime \prime}}{A^5B}=g^{00}G_{tt}+g^{rr}G_{rr}+2g^{\theta\theta}G_{\theta\theta}=0~~~,
\end{equation}
which implies that
\begin{equation}\label{eq:afromb}
A^{\prime\prime}=-\frac{A}{3B} B^{\prime\prime}~~~.
\end{equation}
Combining Eqs. \eqref{eq:einspatial} and \eqref{eq:afromb}, we get the following system of
second order differential equations for $A$ and $B$,
\begin{align}\label{eq:abisoeq}
A^{\prime\prime}=&\frac{3}{4}\Lambda \frac{A^9}{r^4 B^4}~~~,\cr
B^{\prime\prime}=&-\frac{9}{4} \Lambda \frac{A^8}{r^4 B^3}~~~.\cr
\end{align}

Before solving these numerically, we first determine the leading order perturbation corrections in $\Lambda$.
Taking the zeroth order solution to have the standard isotropic
Schwarzschild form
\begin{equation}\label{eq:zerothiso}
A^{(0)}=r+M/2~,~~~B^{(0)}=r-M/2~~~,
\end{equation}
adding order $\Lambda$ corrections by writing
\begin{align}\label{eq:fullab}
A=&A^{(0)}+\Lambda A^{(1)}+...~~~,\cr
B=&B^{(0)}+\Lambda B^{(1)}+...~~~,\cr
\end{align}
substituting into the right hand side of Eq. \eqref{eq:abisoeq}
and integrating twice using Mathematica, we find
\begin{align}\label{eq:firstorder}
A^{(1)\prime}=&\frac{3}{128}\Big[-\frac{M^5}{3 r^3}+\frac{4096 M^5}{3 (M-2
   r)^3}-\frac{13 M^4}{r^2}-\frac{1024 M^4}{(M-2 r)^2}+\frac{4096 M^3}{M-2
   r}-\frac{328 M^3}{r}+2704 M^2 \log (r) \cr
    -&2048 M^2 \log (2 r-M)+208 M r+16
   r^2\Big]~~~,  \cr
B^{(1)\prime}=&-\frac{9}{128} \Big[\frac{M^5}{3 r^3}+\frac{11 M^4}{r^2}-\frac{1024
   M^4}{(2 r-M)^2}+\frac{232 M^3}{r}-1584 M^2 \log (r)+2048 M^2 \log (2
   r-M) \cr
   +&176 M r+16 r^2\Big] ~~~,  \cr
A^{(1)}=&\frac{M^5}{256 r^2}+\frac{8 M^5}{(M-2 r)^2}-\frac{12 M^4}{M-2 r}+\frac{39
   M^4}{128 r}-48 M^3 \log (M-2 r)-\frac{123}{16} M^3 \log (r) \cr
   -&24 M^2 (M-2
   r)-\frac{507 M^2 r}{8}+\frac{507}{8} M^2 r \log (r)-24 M^2 (2 r-M) \log
   (2 r-M)+\frac{39 M r^2}{16}+\frac{r^3}{8}  ~~~, \cr
B^{(1)}=&-\frac{9}{128} \Big[-\frac{M^5}{6 r^2}-\frac{512 M^4}{M-2 r}-\frac{11
   M^4}{r}+232 M^3 \log (r)+1024 M^2 (M-2 r)+1584 M^2 r \cr
   -&1584 M^2 r \log
   (r)+1024 M^2 (2 r-M) \log (2 r-M)+88 M r^2+\frac{16 r^3}{3}\Big]  ~~~. \cr
\end{align}
We have omitted constant of integration terms in the logarithms, since these are
renormalizations of the $r$ and $M$ terms in the zeroth order solution.
For the leading large $r$ terms of $A$ and $B$, these give
\begin{align}\label{eq:firstiso}
A=&r+\frac{M}{2}+\frac{\Lambda}{8} r^3 + O(\Lambda M)~~~,\cr
B=&r-\frac{M}{2}-\frac{3\Lambda}{8} r^3 + O(\Lambda M)~~~;
\end{align}
however, we found that this approximation to Eq. \eqref{eq:firstorder} was
not sufficiently accurate to use to provide initial data for solving the
differential equations.

At this point it is convenient to introduce dimensionless variables
by rescaling $A$, $B$, $r$, and $M$ by a factor of $\Lambda^{1/2}$,
or equivalently, making the replacements $\Lambda \to 1$, $r \to x=\Lambda^{1/2}r$,
$M \to \hat M= \Lambda^{1/2}M$.  This gives as the set of equations to be solved
Eq. \eqref{eq:abisoeq}, with the primes now denoting differentiation with
respect to $x$, and with the leading order solution given by the rescaled
zeroth plus first order solution of Eqs. \eqref{eq:zerothiso} --\eqref{eq:firstorder}.
We use this leading order solution to impose initial conditions for the Mathematica
differential equation integrator at a point $x_0$, which we took for convenience as
$(\hat M)^{1/2}$.  We found that the results were insensitive to changing this value
of $x_0$ by a factor of 10 in either direction (this was not the case when we
tried to use the large $r$ approximation of Eq. \eqref{eq:firstiso} to set the
initial conditions).

We give sample results for a rescaled mass $\hat M=10^{-6}$ in Fig. 5 and Fig. 6.
For this parameter value, the integrator indicated presence of singularities
at $x_{\rm lower}=5.000061 \times 10^{-7}$ and at $x_{\rm upper}=0.8276970$. (In the figures, $x_{\rm lower}$ is
denoted by xmin.) As we changed
$\hat M$, $x_{\rm lower}$ varied as approximately $\hat M/2$ and $x_{\rm upper}$ was approximately
constant to three decimal places.  We interpret the singularity at $x_{\rm upper}$ as
the singularity at $\infty$ found in polar coordinates, and the singularity
at $x_{\rm lower}$ as the singularity  within the black hole.  This interpretation
is supported by two pieces of evidence.  First, when we use
isotropic coordinates to compute the proper distance between the singularities,
\begin{equation}\label{eq:Dcalc1}
D=\Lambda^{-1/2}\int_{x_{\rm lower}}^{x_{\rm upper}} dx A[x]^2/x^2= 0.927374 \Lambda^{-1/2}~~~,
\end{equation}
it is in excellent agreement with the value $D= 0.927371 \Lambda^{-1/2}  $ found in Eq. \eqref{eq:Dcalc}
for the proper distance from $r(polar)=0$ to $r(polar)=\infty$, computed using
the zero mass limit of the $b$ equation.  Second, when we plot $x(polar)=\Lambda^{1/2}r(polar)
=A[x]^2/x$, we see that for $x-x_{\rm lower} > 10^{-10}$ it is a monotone increasing  function of $x$, rising from a very small value
near $x_{\rm lower}$ to a very large value at $x_{\rm upper}$.  Numerically, we found the flat minimum
of $x(polar)$ to be approximately $2 \hat M$, that is, it is $2\times 10^{-6}$ for the plot
of Fig. 6.  This nonzero minimum value, and the rise in $x(polar)$ below $x-x_{\rm lower} = 10^{-10}$, may be reflections of residual inaccuracies in the initial
conditions supplied to the differential equation solver and/or in the numerical integration itself, but this needs
further study.

The plots of Figs 5 and 6 show that there is no horizon between $x_{\rm lower}$ and $x_{\rm upper}$, in agreement with
our polar coordinate results that $g_{00}$ never vanishes, and that the cusp found in polar
coordinates is a coordinate, not a physical singularity.  Thus the
modified effective action static, spherically symmetric Schwarzschild-like solution has very different properties, near
the putative black hole horizon, from the Schwarzschild metric.

\begin{figure}[t]
\centering
\includegraphics{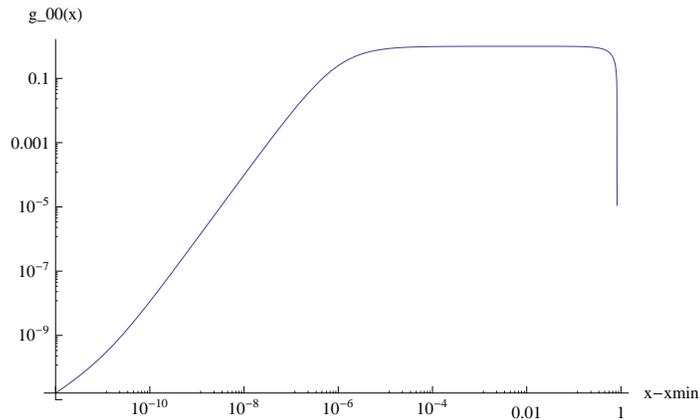}
\caption{Numerical result for $g_{00}(x)$ in isotropic coordinates, versus $x-xmin$,  with $\hat M =M\Lambda^{1/2}=10^{-6}$.
Here $x$ is the dimensionless variable defined by $x=\Lambda^{1/2} r$, with $r$ the radius in isotropic coordinates,
and ${\rm xmin}\equiv x_{lower} \simeq 5 \times 10^{-7}$ is defined in the paragraph containing Eq. \eqref{eq:Dcalc1}.}
\end{figure}

\begin{figure}[t]
\centering
\includegraphics{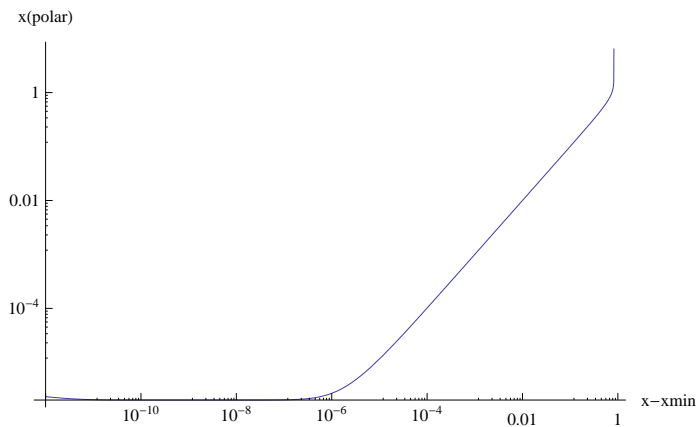}
\caption{Numerical result for $x(polar)\equiv\Lambda^{1/2}r(polar)$ versus $x-xmin$,  with $\hat M =M\Lambda^{1/2}=10^{-6}$.
Here $x$ is the dimensionless variable defined by $x=\Lambda^{1/2} r$, with $r$ the radius in isotropic coordinates; $x(polar)$ and $r(polar)$ are the
corresponding polar coordinate variables defined in Eq. \eqref{eq:dimensionless},
and ${\rm xmin}\equiv x_{lower}\simeq 5 \times 10^{-7}$ is defined in the paragraph containing Eq. \eqref{eq:Dcalc1}.}
\end{figure}
\section{Time dependence}

In the rest of this paper, we will explore whether the bad large distance behavior of the solution of the effective action equations is a
reflection of the restriction to a static solution.  The reason for suspecting this is that when the effective action is used with the standard
Roberston-Walker line element \cite{adlerpaper}, it gives the usual time dependent cosmological solution.  Since the effective action is
invariant only under time-independent coordinate transformations, one cannot proceed by applying a time-dependent coordinate
transformation to the static solution.  Instead, one must study {\it ab initio} time dependent solutions to the effective action equations.

\subsection{Time-dependent generalization of the effective action}

 In turning to time-dependent solutions of the effective action equations, a complication arises, in that when time-dependent metrics are allowed, the form of the effective action changes.  When $g_{0i}=0$, the
general form of the effective action that is invariant under three space general coordinate transformations and the corresponding spatially dependent Weyl transformations employed in the analysis of \cite{adlerpaper}, is
\begin{equation}\label{eq:timedepaction}
\Delta S_{\rm g}=-\frac{\Lambda}{8 \pi G} \int d^4x \frac{(^{(4)}g)^{1/2}}{g_{00}^2} F(\partial_0 g_{00}/g_{00})~~~,
\end{equation}
with $F(u)$ a general function of its argument subject only to the restriction $F(0)=1$.
It is easy to construct examples
\big(such as a Gaussian $F(u)=\exp(-K u^2)$, with $g_{00}(z)=\log(z/a)/\log(z/b)$, $a<b$, $z=r\exp(Ht)$, $\partial_0 g_{00}/g_{00}=H (\log{a/b})/(\log{z/a} \log{z/b})$\big),
to see that the presence of $F$ can make a substantial difference in the nature of
the solutions to the effective action equations.  However, since $F$ is not known, we will not pursue this route, and instead will
look at time-dependent solutions to the effective action equations with $F\equiv 1$. Moreover, we will not study solutions that
involve partial differential equations in $r$ and $t$, but instead focus on a class of time-dependent solutions, first studied
by McVittie \cite{mcvittie}, that reduce to a system of ordinary differential equations in a variable $z=r\exp(Ht)$ that contains both
$r$ and $t$ dependence.

\subsection{Generalized McVittie solution for a de Sitter universe}

We turn now to an examination of time dependent, spherically symmetric solutions to the effective
action equations, focusing on the  Ansatz introduced by McVittie \cite{mcvittie} which leads to
a set of ordinary (rather than partial) differential equations.  In this section we assume
an exponentially expanding de Sitter cosmology; equations (but not numerics) for the general Roberston-Walker cosmology
are given in Appendix C.  The line element for the McVittie Ansatz is
based on isotropic coordinates with an exponential expansion factor included,   in the form
\begin{equation}\label{eq:isomcv}
(ds)^2=\frac{B[r \exp{(Ht)}]^2}{A[r\exp{(Ht)}]^2} (dt)^2-\frac{A[r\exp{(Ht)}]^4 \exp{(-2Ht)}}{r^4} [(dr)^2 + r^2 (d\theta)^2 + \sin^2(\theta) (d\phi)^2)] ~~~,
\end{equation}
with $H$ related to the cosmological constant by
\begin{equation}
3 H^2 = \Lambda~~~.
\end{equation}
We define a composite variable $z$ by
\begin{equation}\label{eq:zdef}
z\equiv r \exp{(Ht)}~~~,
\end{equation}
so that $A$ and $B$ are functions of the argument $z$.
We work with only the $G_{rr}$ and $G_{\theta\theta}$ equations, since as discussed in Appendix B, finding the
conserving extension of $\Delta T^{ij}$ in this case requires solving differential, rather than algebraic,
equations. After some algebraic rearrangement, the $G_{rr}$ and $G_{\theta\theta}$ equations can be put in the form
\begin{align}\label{eq:grrandgthth}
&AB^{\prime}+BA^{\prime}-2zA^{\prime}B^{\prime}=\frac{\Lambda A^9f}{2z^3 B^3}-\frac{H^2 X}{2z^3 B^2}~~~,\cr
&BA^{\prime\prime}+AB^{\prime\prime}=~~~~~~~-3 \frac{\Lambda A^9f}{2z^4 B^3}+3\frac{H^2 X}{2z^4 B^2}~~~,\cr
&X=12 z^2 A^5 B (A^{\prime})^2 + 3 A^7 B + 2 z A^7 B^{\prime}-4 z^2 A^6 A^{\prime}B^{\prime}
 -10 z A^6 B A^{\prime} + 4 z^2 A^6 B A^{\prime\prime}~~~.\cr
\end{align}
The factor $f$ takes the value $f=1$ for the modified effective action of interest here,
and the value $f=g_{00}^2=B^4/A^4$ for a standard cosmological constant action
(the case solved by McVittie with $H^2 =  \Lambda/3$).
From 3 times the first of  Eqs. \eqref{eq:grrandgthth} added to $z$ times the second, we get the simpler equation
\begin{equation}\label{eq:diffmcv}
0=z(B A^{\prime\prime}+A B^{\prime\prime}) + 3(A B^{\prime}+B A^{\prime}-2z A^{\prime}B^{\prime})~~~.
\end{equation}
The first equation of Eqs. \eqref{eq:grrandgthth} can be solved for $A^{\prime\prime}$ in terms
of $A$, $B$, $A^{\prime}$, $B^{\prime}$, and Eq. \eqref{eq:diffmcv} can then be used to get a
similar equation for $B^{\prime\prime}$, which can be used as input to a differential equation solver.

Before turning to numerical solutions, we first calculate analytic approximations to the equation system
in the small $z$ and large $z$ regimes.  For small $z$ we make a perturbation expansion in the parameters
$\Lambda$ and $H^2$.  Writing $H^2 = \xi \Lambda$, the perturbation parameter is $\Lambda$.  Taking as
the zeroth order solution $A^{(0)}=z+M/2$~,~~ $B^{(0)}=z-M/2$, we write
\begin{align}\label{eq:pertdefs}
A=&z+M/2+\Lambda A^{(1)}~~,~~~
B=z-M/2+\Lambda B^{(1)}~~~,\cr
A^{\prime}=&1+\Lambda A^{(1)\prime}~~,~~~
B^{\prime}=1+\Lambda B^{(1)\prime}~~~,\cr
A^{\prime\prime}=&\Lambda A^{(1)\prime\prime}~~,~~~
B^{\prime\prime}=\Lambda B^{(1)\prime\prime}~~~.\cr
\end{align}

The left hand sides of Eqs. \eqref{eq:grrandgthth} become, to first order in $\Lambda$,
\begin{align}\label{eq:grrgththfirst}
&AB^{\prime}+BA^{\prime}-2zA^{\prime}B^{\prime} \simeq \Lambda[A^{(1)} + B^{(1)}- (A A^{(1)\prime} + B B^{(1)\prime}]~~~,\cr
&BA^{\prime\prime}+AB^{\prime\prime}\simeq \Lambda[ B A^{(1)\prime\prime} + A B^{(1)\prime\prime}]~~~.\cr
\end{align}
Diffentiating the right hand side of the first line of this equation with respect to $z$, it becomes
\begin{equation}\label{eq:grrgththfirst1}
(d/dz)\Lambda[A^{(1)} + B^{(1)}- (A A^{(1)\prime} + B B^{(1)\prime})]\simeq -\Lambda(A A^{(1)\prime\prime} + B B^{(1)\prime\prime})~~~,
\end{equation}
giving an independent linear combination of $A^{(1)\prime\prime}$ and $B^{(1)\prime\prime}$ from that appearing in
the second line.
Combining  Eqs. \eqref{eq:grrgththfirst} and \eqref{eq:grrgththfirst1} with Eq. \eqref{eq:grrandgthth},
factoring away $\Lambda$, and solving for   $A^{(1)\prime\prime}$ and $B^{(1)\prime\prime}$, we get as the perturbation equations
\begin{align}\label{eq:finalpert}
A^{(1)\prime\prime}=&\frac{3}{4}\frac{A^9}{z^4B^4}+\frac{\xi}{B^2-A^2}\left[\frac{3}{2}\frac{X}{z^4B}-A \frac{d}{dz}\left(\frac{X}{2z^3B^2}\right)\right]~~~,\cr
B^{(1)\prime\prime}=&-\frac{9}{4}\frac{A^8}{z^4B^3}+\frac{\xi}{B^2-A^2}\left[B \frac{d}{dz}\left(\frac{X}{2z^3B^2}\right)-\frac{3}{2}\frac{AX}{z^4B^2}\right]~~~,\cr
\end{align}
with $A$ and $B$ on the right the leading order solutions $A^{(0)}$ and $B^{(0)}$ given above, and with $X$ on the right the leading order part $X^{(0)}=3(A^{(0)})^5 (B^{(0)})^3$.  Carrying out the differentiation in the $\xi$ term and simplifying, we get as the equations to be integrated
\begin{align}\label{eq:finalpert2}
A^{(1)\prime\prime}=&\frac{3}{4}\frac{A^9}{z^4B^4}+\frac{3 \xi A^5}{4 z^4}~~~,\cr
B^{(1)\prime\prime}=&-\frac{9}{4}\frac{A^8}{z^4B^3}+\frac{15 \xi A^4 B} {4 z^4}~~~,\cr
\end{align}
again with $A$ and $B$ on the right the leading order solutions.
The terms on the right that do not have a factor $\xi$ are just
the perturbations of the static isotropic equations of Eq. \eqref{eq:abisoeq}, while the terms proportional to $\xi$ are the additional terms
resulting from the time dependence of the generalized McVittie Ansatz of Eq. \eqref{eq:isomcv}.  Setting $\xi=1/3$, the value which makes
$A=B=z$ an exact solution of Eqs. \eqref{eq:grrandgthth}, and integrating Eqs. \eqref{eq:finalpert} using Mathematica, we find (again omitting
constant of integration terms in the logarithms)
\begin{align}\label{eq:finalpert1}
A^{(1)\prime}=&  \frac{1}{128} \Big(-\frac{4 M^5}{3 z^3}+\frac{4096 M^5}{(M-2
   z)^3}-\frac{44 M^4}{z^2}-\frac{3072 M^4}{(M-2 z)^2}+\frac{12288
   M^3}{M-2 z}-\frac{1024 M^3}{z}+8192 M^2 \log (z)\cr-&6144 M^2 \log (2
   z-M)+704 M z+64 z^2\Big)  ~~~,\cr
A^{(1)}=&\frac{M^5}{192 z^2}+\frac{8 M^5}{(M-2 z)^2}-\frac{12 M^4}{M-2 z}+\frac{11
   M^4}{32 z}-48 M^3 \log (M-2 z)-8 M^3 \log (z)-24 M^2 (M-2 z)\cr-&64 M^2
   z+64 M^2 z \log (z)-24 M^2 (2 z-M) \log (2 z-M)+\frac{11 M
   z^2}{4}+\frac{z^3}{6}    ~~~,\cr
B^{(1)\prime}=&  \frac{1}{128} \Big(-\frac{4 M^5}{3 z^3}-\frac{84 M^4}{z^2}+\frac{9216
   M^4}{(2 z-M)^2}-\frac{2048 M^3}{z}+14336 M^2 \log (z)-18432 M^2 \log (2
   z-M)\cr-&1344 M z-64 z^2\Big)  ~~~,\cr
B^{(1)}=& \frac{M^5}{192 z^2}+\frac{36 M^4}{M-2 z}+\frac{21 M^4}{32 z}-16 M^3 \log
   (z)-72 M^2 (M-2 z)-112 M^2 z+112 M^2 z \log (z)\cr-&72 M^2 (2 z-M) \log (2
   z-M)-\frac{21 M z^2}{4}-\frac{z^3}{6}   ~~~.\cr
\end{align}
We note for future reference that the leading terms at large $z$ of $A^{(1)}$ and $B^{(1)}$ are respectively
$z^3/6$ and $-z^3/6$.

We next give a second way of approximating Eqs. \eqref{eq:grrandgthth}, which we now rewrite as Eq. \eqref{eq:diffmcv},
together with a second independent linear combination, where we have set $H^2=\Lambda/3$,
\begin{align}\label{eq:Seq}
&(AB^{\prime}+B A^{\prime})-2z A^{\prime}B^{\prime}=\frac{-\Lambda S}{2z^3B^3}~~~,\cr
S=&2/3) z^2A^6 B (B A^{\prime\prime}-A B^{\prime\prime})-A^9  + A^7B^2+4z^2A^5B^2(A^{\prime})^2\cr
-&(4/3) z A^7 B B^{\prime} +(8/3)z^2 A^6 B A^{\prime}B^{\prime}-(16/3)z A^6 B^2 A^{\prime}~~~.\cr
\end{align}
As in our handling of the static isotropic case, we again introduce dimensionless variables
by rescaling $A$, $B$, $r$, and $M$ by a factor of $\Lambda^{1/2}$,
or equivalently, making the replacements $\Lambda \to 1$, $r \to x=\Lambda^{1/2}r$,
$M \to \hat M= \Lambda^{1/2}M$. This leaves $\hat M$ as the small parameter in the
problem, so we look for a perturbation solution around the Roberston-Walker
cosmological solution $A=B=x$ as modified by the presence of a point mass.
Thus we now make a perturbation expansion in the form
\begin{align}\label{eq:xmpert}
A=&x+\hat Mf_A~~,~~~
B=x+\hat Mf_B~~~,\cr
A^{\prime}=&1+\hat Mf_A^{\prime}~~,~~~
B^{\prime}=1+\hat Mf_B^{\prime}~~~,\cr
A^{\prime\prime}=&\hat Mf_A^{\prime\prime}~~,~~~
B^{\prime\prime}=\hat Mf_B^{\prime\prime}~~~.\cr
\end{align}
It will be convenient to define sum and difference perturbations
\begin{equation}\label{eq:sumpert}
\sigma=f_A+f_B~,~~~\delta=f_A-f_B~~~.
\end{equation}
From Eq. \eqref{eq:diffmcv} we find
\begin{equation}\label{eq:sigmaeq}
0=x^2\sigma^{\prime\prime}+3(\sigma-x \sigma^{\prime})~~~,
\end{equation}
which has the general solution
\begin{equation}\label{eq:sigmasoln}
\sigma=c_1 x + c_2 x^3~~~,
\end{equation}
and from Eq. \eqref{eq:Seq} we find
\begin{equation}\label{eq:deltaeq}
(1/3) x^4 \delta^{\prime\prime} + x^2(-2\delta+x\delta^{\prime})=(-\sigma+x\sigma^{\prime})[1-(5/3) x^2]~~~,
\end{equation}
which has the general solution
\begin{align}\label{eq:deltasoln}
\delta=&c_3 x^{\lambda_+} +c_4 x^{\lambda_-}-c_2[2x+(10/9) x^3]~~~,\cr
\lambda_+=&-1+\surd 7~,~~~\lambda_-=-1-\surd 7~~~,\cr
\end{align}
with $c_1$ through $c_4$ numerical constants that can depend on  $\hat M$ through matching to the solution for $x <<1$.

From these solutions we learn two interesting facts.  First, neglecting the term $\hat M c_1 x$ (which is a renormalization of the leading solution $x$), for large $x$ the metric component $g_{00}$ behaves
as
\begin{equation}\label{eq:g00largex}
g_{00}=\frac{B^2}{A^2}=\left[  \frac{x+\hat M[(19/18)c_2 x^3-(1/2) c_3 x^{\lambda_+}]}{ x+\hat M[-(1/18)c_2 x^3+(1/2) c_3 x^{\lambda_+}] } \right]^2~~~.
\end{equation}
Depending on the magnitudes and signs of $c_2$ and $c_3$, either $B$ vanishes or $A$ vanishes for large enough $x$, that is, either $g_{00}$ vanishes or $g_{00}$ becomes singular.  If the $x^3$ term dominates, this will occur at
\begin{equation}\label{eq:xcut1}
x_{\rm cutoff}\sim \big(1/(\hat M c_2)\big)^{1/2} ~~~,
\end{equation}
while if the $x^{\lambda_+}$ term dominates, this will occur at
\begin{equation}\label{eq:xcut2}
x_{\rm cutoff}\sim \big(1/(\hat M c_3)\big)^{1/(\surd 7 -2)} ~~~.
\end{equation}
Thus in either case, as $\hat M$ approaches zero, provided $\hat M c_2$ and $\hat M c_3$ also approach zero, the large-$x$ singularity found in the static case runs off to infinity, when time dependence is allowed through the McVittie Ansatz. Since values of $\hat M$ of astrophysical interest are very small, the problem found in the static case can move off to super-cosmological distances $x>>1$.

Second, we note that the leading terms of the $\Lambda$ perturbations, which we saw were $\pm \Lambda z^3/6$,
after rescaling by $\Lambda^{1/2}$ become $\pm \Lambda^{3/2} z^3/6=\pm x^3/6$, which do not match the
leading $x^3$ terms $ \hat M c_2 x^3$  in Eqs. \eqref{eq:sigmasoln} and \eqref{eq:deltasoln}.  Hence
if there is a continuous solution linking the small-$x$ and large $x$  regions, in between them  there must be a transitional region, where the $x^3$ term
strongly decreases in magnitude. Since we do not have an analytic approximation to the behavior in this region,
in using the $\Lambda$ perturbation to furnish initial data at a point $x_0$, we must take care to choose
$x_0<<1$.  We must also take $x_0>>\hat M$ since the perturbation expansion becomes singular at $x=2\hat M$.
In sum, the point $x_0$ at which perturbative initial data is imposed must lie in the range $\hat M << x_0 << 1$.
(An alternative possibility to having a continuous solution linking the small-$x$ and large $x$  regions, which we will discuss further in the next section, is that all solutions starting in the small-$x$ region become
singular before the large $x$  region is reached.)

\subsection{Numerical results for the McVittie Ansatz}

We turn now to numerical results for the generalized McVittie solution for the de Sitter universe.  We find the
numerical solution to the McVittie equations to be very sensitive to truncation errors, and beyond this, to the choice of initial conditions and integration parameters.   To minimize the effect of truncation errors, we rewrite the
equations for $A$ and $B$ in terms of new dependent variables $f$ and $g$ (which up to a factor of
$\hat M$ are the same as the variables $\sigma$ and $\delta$ used in the analysis of Eqs. \eqref{eq:xmpert} --
\eqref{eq:deltasoln}),
\begin{align}\label{eq:fandg}
A=&z+(f+g)/2~~~,\cr
B=&z+(f-g)/2~~~.
\end{align}
Scaling $\Lambda$ out of the equations by setting $\Lambda=1$ and setting $H^2/\Lambda =1/3$, we then algebraically cancel out the numerator terms that contain only $z$; this is possible because as noted above, for $H^2=\Lambda/3$, we know that $A=B=z$ is an exact solution of the equations.  When the first line of Eq. \eqref{eq:grrandgthth} is solved for
$A^{\prime\prime}$, the term most sensitive to truncation errors is
\begin{equation}\label{eq:sensterm}
\frac{3 z B}{2 A^6} [A B^{\prime}+BA^{\prime}-2zA^{\prime}B^{\prime}]~~~.
\end{equation}
In terms of the $f$ and $g$ variables, the square bracket in this equation becomes
\begin{equation}\label{eq:sensterm1}
f - z f^{\prime} + [f f^{\prime} -g g^{\prime} +z \big((g^{\prime})^2-(f^{\prime})^2\big)]/2~~~,
\end{equation}
which is second order in small quantities except for the first order terms $f- z f^{\prime}$.
However, the  order $(0)$  contribution to $f$ is zero, and in the order $(1)$
contributions the $\hat M$-independent terms $\pm z^3/3$ and $\pm z^2/2$ cancel.
As a result the first order terms $f -z f^{\prime}$ are also very small, and so the change of variables
of Eq. \eqref{eq:fandg} succeeds in reducing truncation errors.    The algebraic
rearrangement of the rest of Eq.  \eqref{eq:grrandgthth} is complicated enough
that it had to be done by Mathematica, so we won't write down the explicit formulas.

Even with minimization of truncation errors, the solution of the generalized McVittie equations
is very sensitive to the point at which the initial values are imposed, and to  integration parameters
that control the distribution of integration points. We do not know whether this is because the system of
equations exhibits chaotic behavior, or because the Mathematica integrator is giving deceptive
results.   We find two general classes of solutions: (i)
solutions that become singular at values $0.001<x_{\rm cutoff}<0.1$ of the rescaled variable $x$,  which lie
in the transitional region between the small $x$  and large $x$  regions discussed in the previous section,
and (ii) solutions that are well-behaved all the way from the small $x$  region, through the transitional region,
out to values $x_{\rm cutoff}>10$ which
lie in the large $x$  region.   Even though we cannot establish that the latter solutions are true solutions of
the differential equations, and not an artifact of the Mathematica integrator, for completeness we give results from a typical solution of type (ii).

For this solution, we choose the
same mass parameter as in the static isotropic case, $\hat M=10^{-6}$,  and apply the zeroth plus first order
$\Lambda$ perturbation series as initial conditions at $x_0=0.55 \times10^{-3}$ (a point of somewhat
greater stability than $x_0=10^{-3}$).  We use the Mathematica integrator NIntegrate with Precision and
Accuracy goals 6, and with the maximum of the integration range $xmax=90$.  With these choices,  the qualitative results are stable for $x_0$ in the range from $x_0=0.50 \times 10^{-3}$ to $x_0=0.60 \times 10^{-3}$, but the location $x_{\rm cutoff}$ of the large $x$  singularity is highly sensitive to the choice of $x_0$ (varying from $\sim 26$ to $\sim89$  over this range).  The results are also sensitive to variation of $xmax$, which affects the distribution of integration points.  The qualitative results are stable for $xmax$ in the range 85 to 95, again with the location $x_{\rm cutoff}$ of the large $x$  singularity highly sensitive to the choice of $xmax$ (varying from $\sim 11$ to $\sim 89$  over this range).
As long as the integration minimum  is smaller than $x_0$, the results for $x>x_0$ are not sensitive to where
the minimum is placed. So effectively, the solutions are governed by two continuum parameters $xmax$ and $x_0$,
and it would be interesting to find a way to map out the distribution of type-(i) and type-(ii) solutions in the
$xmax$--$x_0$ plane.  For the particular solution shown in the graphs, $x_{\rm cutoff} \simeq 89.3$, and the
proper distance integral ie $\int_{\hat M/2}^{x_{\rm cutoff}} dx A[x]^2/x^2 \simeq 92.2 $.

In Figs. 7 and 8 we plot comparisons of the numerical solution with fits to the large $x$  analysis of  Eqs. \eqref{eq:xmpert} --\eqref{eq:deltasoln}.  The fits were done by least squares over the interval
(0.1,10) with a $1/x$ weighting, that is, an integration  measure $\int_{0.1}^{10} d{\rm log}x=\int_{0.1}^{10} dx/x$, corresponding to use of a logarithmic $x$ axis in the figures.   One gets for the coefficients in
Eqs. \eqref{eq:xmpert} -- \eqref{eq:deltasoln},
\begin{align}\label{eq:fitcoeff}
c1=&-4.85557 \times 10^{-6}~~~,\cr
c2=&-0.000094006 ~~~,\cr
c3=&-0.00153208 ~~~,\cr
c4=&-1.09393 \times 10^{-10}~~~.\cr
\end{align}
In Fig. 7 we plot $d(A+B) \equiv 2x-(A[x]+B[x])$ from the numerical solution (solid line) as compared with $-(c1\, x + c2\, x^3)$ (dashed line).  In Fig. 8 we plot $d(A-B)\equiv -(A[x]-B[x]+c2(2x + (10/9) x^3))$ from the numerical
solution (solid line) as compared with $-(c3\, x^{\lambda_{+}} + c4 \,x^{\lambda_{-}})$ (dashed line).
(In these, and subsequent plots,
the absicssa is $x-xmin$, with $xmin=5\times 10^{-7}= \hat M/2$, corresponding to the choice made in the static
istropic plots.)  The good agreement of the large $x$  computational results with the analytic large $x$  prediction
indicates that the differential equation solver is giving reliable results for $x>0.1$, but this does not rule out the
possibility that the integrator is jumping over a singularity lying below $x=0.1$.

In  Figs. 9, 10, and 11 we give plots that can be compared with Figs. 5 and 6 of the static isotropic case.
In Fig. 9  we plot $g_{00}$ out to large $x$, showing that the singularity has moved from $ \sim 0.83$
of the static isotropic case to $ \sim 89$ when time dependence is included.  In Fig. 10 we plot
$g_{00}$ out to $x=1$, for comparison with the static isotropic case shown in Fig. 5, and in Fig. 11 we plot
the polar coordinate radius $x(polar)=A[x]^2/x$ for comparison with Fig. 6 of the static isotropic case.
The resemblance is striking, indicating that at small $x$ the generalized McVittie solution, if not
an integrator artifact, has a
similar structure to the Schwarzschild-like solution studied in the static case.

We conclude that (i) introducing time-dependence significantly changes the static solutions studied earlier, and (ii)
subject to the caveats discussed above, there may be
smooth solutions that interpolate between a Schwarzschild-like solution at
short distances, and a solution that has $g_{00}=1$ out to super-cosmological distances.

\begin{figure}[t]
\centering
\includegraphics{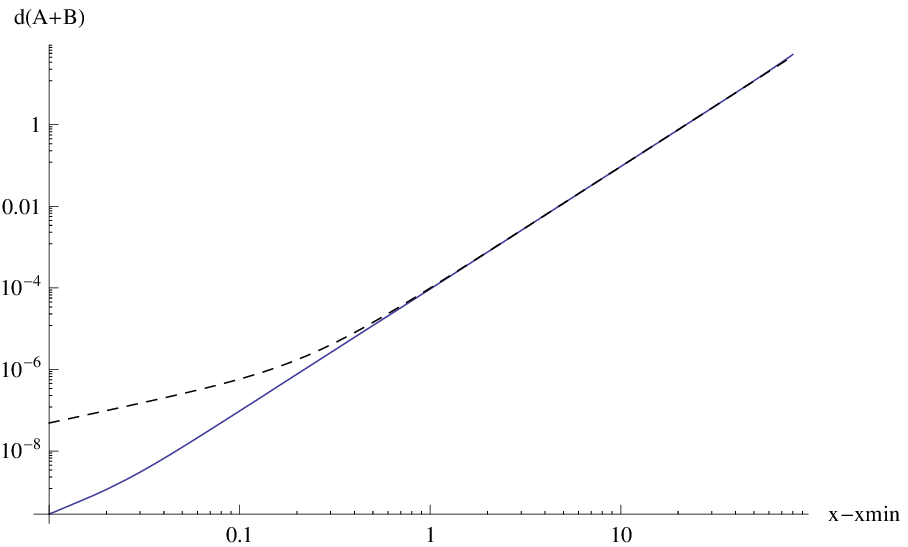}
\caption{Plot of $d(A+B)$ (defined in text) for the numerical solution of the McVittie Ansatz equation (solid line) compared to  a fit to the large $x$  analytic
approximation (dashed line),  versus $x-xmin$, with  $x$ the dimensionless variable defined  following Eq. \eqref{eq:Seq}, and ${\rm xmin}=5\times 10^{-7}$. }
\end{figure}

\begin{figure}[t]
\centering
\includegraphics{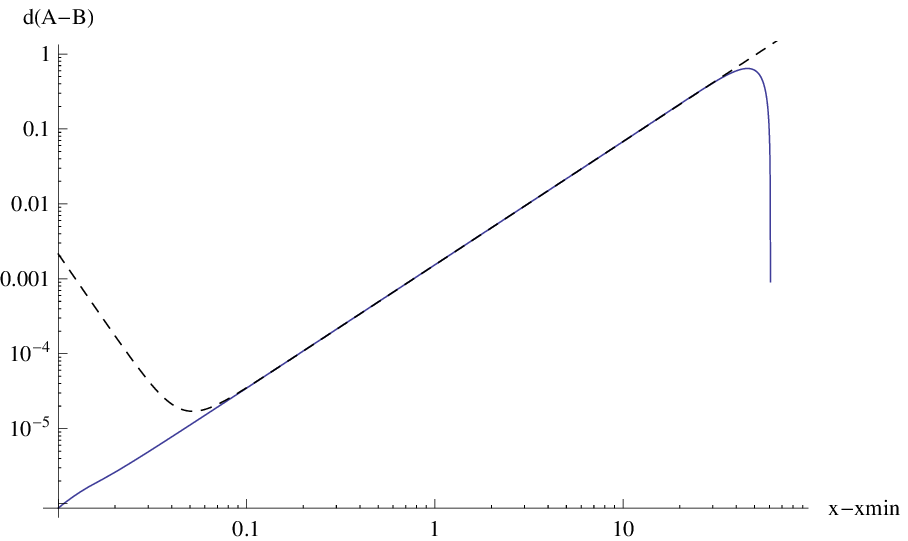}
\caption{Plot of $d(A-B)$ (defined in text) for the numerical solution of the McVittie Ansatz equation (solid line) compared to a fit to the large $x$  analytic
approximation (dashed line), versus $x-xmin$, with  $x$ the dimensionless variable defined  following Eq. \eqref{eq:Seq}, and ${\rm xmin}=5\times 10^{-7}$. }
\end{figure}

\begin{figure}[t]
\centering
\includegraphics{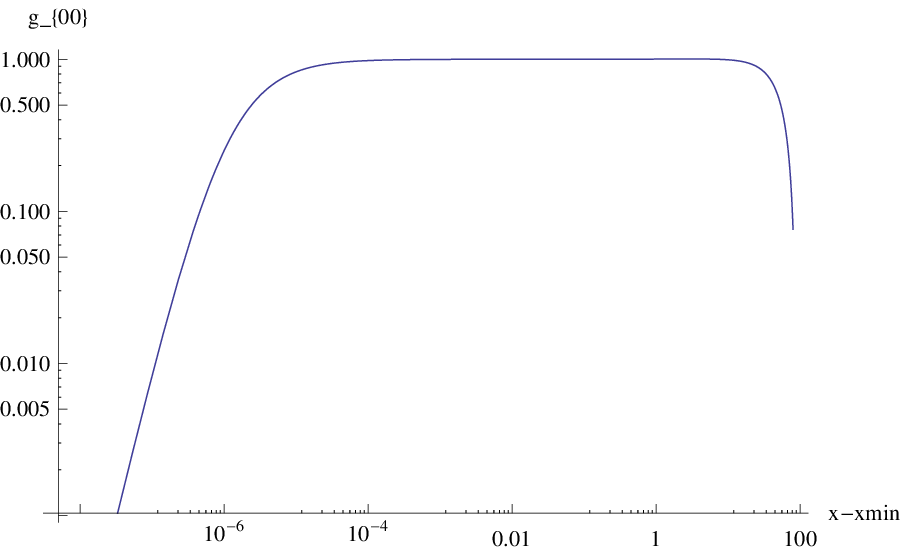}
\caption{Numerical result for $g_{00}(x)$ for the numerical solution of the McVittie Ansatz equation versus $x-xmin$,  with $\hat M =M\Lambda^{1/2}=10^{-6}$. Here $x$ is the dimensionless variable defined  following Eq. \eqref{eq:Seq}, and ${\rm xmin}=5\times 10^{-7}$. }
\end{figure}

\begin{figure}[t]
\centering
\includegraphics{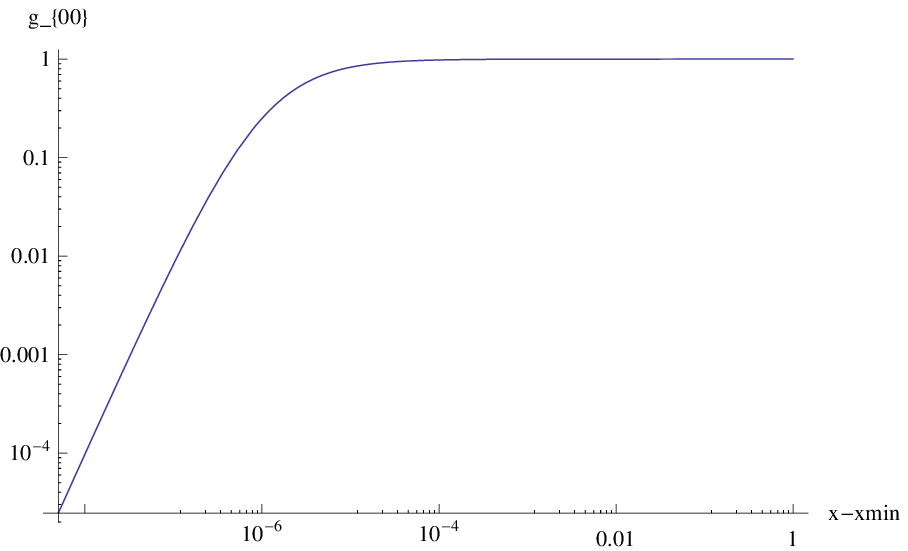}
\caption{Numerical result for $g_{00}(x)$ for the numerical solution of the McVittie Ansatz equation versus $x-xmin$,  with $\hat M =M\Lambda^{1/2}=10^{-6}$. Here $x$ is the dimensionless variable defined  following Eq. \eqref{eq:Seq}, and ${\rm xmin}=5\times 10^{-7}$. }
\end{figure}

\begin{figure}[]
\centering
\includegraphics{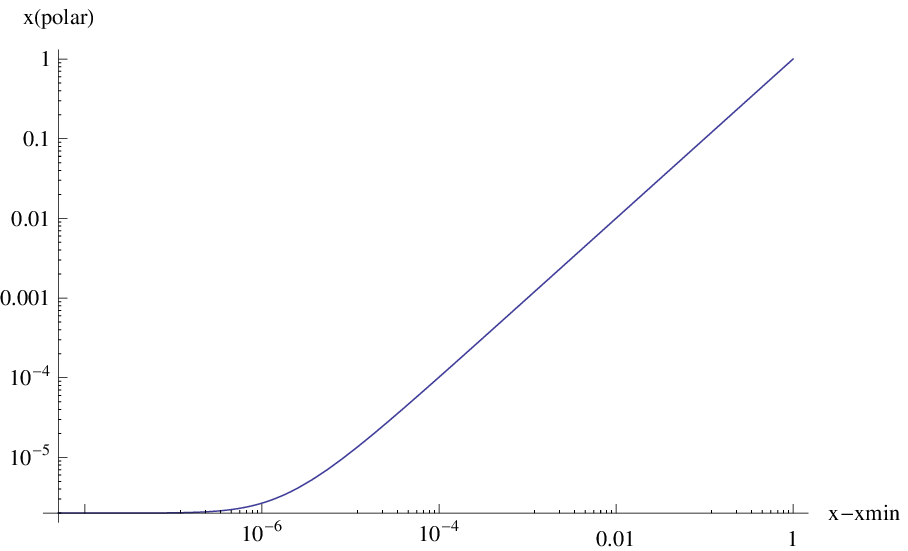}
\caption{Numerical result for $x(polar)=A[x]^2/x$ for the numerical solution of the McVittie Ansatz equation versus $x-xmin$,  with $\hat M =M\Lambda^{1/2}=10^{-6}$. Here $x$ is the dimensionless variable defined  following Eq. \eqref{eq:Seq}, and ${\rm xmin}=5\times 10^{-7}$.}
\end{figure}

\section{Conclusion and Discussion}

We have studied in detail the modifications to the Schwarzschild black hole solution of general relativity resulting from
 adding the effective action of Eq. \eqref{eq:action2} to the usual Einstein-Hilbert action. Our results can be summarized as
 follows:

 (1)  For radii that are more than $10^{-17} {\cal M}$ cm
 from the Schwarschild radius for a black hole with mass $\cal M$ in solar mass units, but are much less than cosmological distances of order $H_0^{-1}$ (with $H_0$ the Hubble
 constant), the solution is very similar to the usual Schwarzschild solution.  Hence the astrophysics of black holes at the
 core of galaxies and their accretion disks is basically unchanged.

 (2)  Within $10^{-17}{\cal M}$ cm of the Schwarzschild radius, the solution is radically altered.  There is no horizon where
 $g_{00}$ vanishes, and the solution in isotropic coordinates is smooth until the central singularity is reached.  Thus the black hole
 structure near the Schwarzschild radius, when the cosmological constant is introduced through the Lorentz noninvariant action of Eq. \eqref{eq:action2}, is in marked contrast to the black hole horizon found  when the cosmological constant is introduced in the usual way through a Lorentz invariant
 action term.  As we have seen,  the latter leads to the Schwarzschild-de-Sitter solution discussed in Sec. IIB, which has a black hole horizon essentially the same as when
 the cosmological constant is zero.

 (3) At cosmological proper distances of order $H_0^{-1}$ there is another singularity.  Like the central singularity,
 this is a physical singularity, not a coordinate singularity, since curvature invariants become infinite there.  But
 it may be an artifact of the assumption of a static, time-independent solution, since the form of the solutions is greatly altered when time dependence
 is allowed.

 The absence of a horizon implies that
  our solution does not obey the
 ``cosmic censorship'' hypothesis \cite{penrose}  which states that singularities should be screened by horizons, that is,
 there are no ``naked singularities''.  However, whether the central singularity of our solution should be called a ``naked singularity'' is a matter of debate, since this term is usually
 applied to solutions of the classical Einstein equations without quantum or pre-quantum corrections.  Moreover,
 when the solutions of the modified Einstein equations studied in this paper become singular or even rapidly varying,
use of the effective action of Eqs. \eqref{eq:action1}--\eqref{eq:action2}, in which derivatives
 of the metric are neglected,  is no longer justified.  Hence our analysis leads to no firm predictions about the
 nature of the central singularity, or indeed, whether there is one in the exact theory.

 The most important implication of the absence of a horizon in our modified solution is that the ongoing debate about
 black holes and quantum information loss \cite{giddings},  black hole firewalls \cite{almheiri}, and whether there is or is not a true
 horizon in quantum black holes \cite{barbieri, chapline}, will be altered.  But at the current stage it would be premature
 to attempt to say more, since  this will require a further detailed investigation,
 to which the studies of this paper are a necessary preliminary.

\section{Acknowledgements}
 This work was supported in part by the National Science Foundation under
Grant No. PHYS-1066293 and the hospitality of the Aspen Center for Physics.

\appendix

\section{Notational conventions}

Since many different notational conventions are in use for gravitation and cosmology, we summarize here
the notational conventions used in this paper and in \cite{adlerpaper} .

(1)~~ The Lagrangian in flat spacetime is $L=T-V$, with $T$ the kinetic energy and $V$ the potential energy, and
the flat spacetime Hamiltonian is $H=T+V$.

(2)~~ We use a $(1,-1,-1,-1)$ metric convention, so that in flat spacetime, where the metric is denoted by $\eta_{\mu \nu}$, the various 00 components of
the stress energy tensor $T_{\mu \nu}$ are equal, $T_{00}=T_{0}^{0}=T^{00}$.

(3)~~The affine connection, curvature tensor, contracted curvatures, and the Einstein tensor, are
given by
\begin{align}
\Gamma^{\lambda}_{\mu \nu} =&\frac{1}{2}g^{\lambda \sigma}(g_{\sigma \nu,\, \mu}+g_{\sigma \mu,\, \nu}- g_{\mu \nu,\, \sigma})~~~,\cr
R^{\lambda}_{\tau \mu \nu}=& \Gamma^{\lambda}_{~\tau \mu,\, \nu}-\Gamma^{\lambda}_{~\tau \nu,\, \mu} +{\rm quadratic ~terms ~in~} \Gamma  ~~~,\cr
R_{\mu \nu}=&R^{\lambda}_{~\mu \lambda \nu}=-\Gamma^{\lambda}_{\mu \nu, \, \lambda}+{\rm other~terms}~~~,\cr
R=&g^{\mu \nu} R_{\mu \nu}~~~,\cr
G_{\mu \nu}=&R_{\mu \nu}-\frac{1}{2} g_{\mu \nu} R ~~~.\cr
\end{align}

(4)~~ The gravitational action, with {\it bare} cosmological constant $\Lambda_0$, and its variation with respect to the metric $g_{\mu \nu}$ are
\begin{align}\label{eq:lambda0}
S_{\rm g} = & \frac{1}{16\pi G} \int d^4x (^{(4)}g)^{1/2} (R-2\Lambda_0)~~~,\cr
\delta S_{\rm g}=& -\frac{1 }{16 \pi G} \int d^4x (^{(4)}g)^{1/2} (G^{\mu \nu}+\Lambda_0 g^{\mu \nu})\delta g_{\mu \nu}~~~.\cr
\end{align}

(5)~~The matter action and its variation with respect to the metric $g_{\mu \nu}$ are
\begin{align}\label{eq:matteraction}
S_{\rm m}=& \int dt  L = \int d^4 x (^{(4)}g)^{1/2}  {\cal L}(x)~~~,\cr
\delta S_{\rm m}=&-\frac{1}{2} \int d^4 x (^{(4)}g)^{1/2}  T^{\mu \nu}  \delta g_{\mu \nu}~~~.\cr
\end{align}

(6)~~The Einstein equations are
\begin{equation}
G^{\mu \nu}+ \Lambda_0 g^{\mu \nu} +8 \pi G T^{\mu \nu}=0~~~.
\end{equation}

Throughout this paper (except in  Sec. IIB) , we take the bare cosmological constant $\Lambda_0$ to be zero, and instead include an
induced cosmological constant through the frame dependent effective action of Eqs.  \eqref{eq:action1} and \eqref{eq:action2}.
For a Robertson-Walker cosmological metric, this effective action term behaves like a bare cosmological constant term, but for other
metrics it has very different implications.
\section{Conserving extension calculations}

We first give a quick way of finding the conserving extension
extension for $\Delta T^{ij}$ of Eq. \eqref{eq:tijeqs}.  The problem
is to find a factor $F$ such that $\Delta T^{00}=(\Lambda/(8\pi G))F g^{00}$ obeys
the covariant conservation equation
\begin{equation}\label{eq:tcovcons}
\nabla_{\mu} \Delta T^{\mu i}=0~~~.
\end{equation}
Writing $F=F-1/g_{00}^2 +1/g_{00}^2$, and using the fact that $\nabla_{\mu} g^{\mu i}=0$,
for a diagonal metric this becomes
\begin{equation}\label{eq:deriv00}
\partial_i (1/g_{00}^2) g^{ii} + \Gamma_{00}^i g^{00} (F-1/g_{00}^2)=0~~~,
\end{equation}
with $\Gamma^{\lambda}_{\mu \nu}$ the affine connection and no sum on $i$, giving an algebraic equation for $F$.  Since
for a diagonal, static metric $\Gamma_{00}^i = -(1/2)g^{ii}\partial_i g_{00}$, again with
no sum on $i$, Eq. \eqref{eq:deriv00} becomes
\begin{equation}
\partial_i g_{00}g^{ii}[-2 g_{00}^{-3} -(1/2)g^{00} (F-1/g_{00}^2)]=0~~~.
\end{equation}
This equation has two branches.  When $\partial_i g_{00} \neq 0$,
it requires
\begin{equation}
F=-3/g_{00}^2~~,~~~\Delta T_{00}=(\Lambda/(8\pi G))F g_{00}=(\Lambda/(8\pi G))(-3/ g_{00})~~~,
\end{equation}
as used in the text.  There is also a solution with $\partial_i g_{00}=0$, corresponding
to the cosmological Robertson-Walker case of $g_{00}=1~,~~~\Delta T^{\mu\nu}=(\Lambda/(8\pi G)) g^{\mu\nu}$
discussed in \cite{adlerpaper}.

We turn next to the conserving extension for the generalized McVittie solution studied in Sec. IVB.  Although
the metric in this case is still diagonal ($g_{tr}=0$), calculating the Einstein tensor shows that the component
$G_{tr}\neq 0$.  Hence the conserving extension of $\Delta T_{rr}(z)$ and $\Delta T_{\theta \theta}(z)$ will have
components $\Delta T_{tt}(z)$ and $\Delta T_{tr}(z)=\Delta T_{rt}(z)$, with $z=r \exp{(Ht)}$.  Working out the
covariant conservation equations, one finds two linear differential equations in the unknowns $\Delta T_{tt}(z)$
and $\Delta T_{tr}(z)$ and their first derivatives with respect to $z$, with coefficients constructed from $z$ and from
$A(z)$ and $B(z)$ and their derivatives. These equations can in principle be integrated once $A(z)$ and
$B(z)$ are known, but one can no longer solve for the conserving extension algebraically. Hence in Sec. IVB we
work exclusively with the $G_{rr}$ and $G_{\theta\theta}$ equations, which suffice to determine $A(z)$ and $B(z)$.
Once these are known, calculating the corresponding Einstein tensor components (which automatically obey the
Bianchi identities) gives directly the conserving
extensions $\Delta T_{tt}(z)$ and $\Delta T_{tr}(z)$ that obey the covariant conservation equations.
Since the covariant conservation equations do not factorize into two disjoint branches, the McVittie solution can allow
a crossover from a static-like solution at small $z$ to a cosmological-like solution at large (but still finite) $z$,
as discussed in Secs. IVB and IVC.

\section{Generalized McVittie equations for a general expansion factor}

The Ansatz of Eq. \eqref{eq:isomcv} can be extended to the case when the expansion factor $\exp{(Ht)}$ is replaced by by a general
$R(t)$.  The composite variable $z$ is now defined by
\begin{equation}\label{eq:zdef1}
z\equiv r R(t)~~~,
\end{equation}
and the differential equations are changed as follows.
In Eq. \eqref{eq:grrandgthth}, $H^2 X$ is replaced by
\begin{align}
H^2 X \to& \left( \frac{\dot R}{R}\right)^2  X + \left[ \left( \frac{\dot  R}{R}\right)^2-\frac{\ddot R}{R}\right] Y~~~,\cr
Y =&2A^7B-4zA^6BA^{\prime} = 2A^6B(A-2zA^{\prime})~~~,\cr
\end{align}
with the dot denoting differentiation of $R(t)$ with respect to $t$,
while Eq. \eqref{eq:diffmcv} is unchanged.
From the Friedmann equations for a spatially flat ($k=0$) matter dominated universe, the expressions involving
$R(t)$ after scaling out $\Lambda$ can be evaluated  as
\begin{align}\label{eq:fried}
\xi \equiv &\Lambda^{-1}\left(\frac{\dot R}{R}\right)^2= \frac{1}{3\Omega_{\Lambda}}~~~,\cr
\eta \equiv &\Lambda^{-1}\left[ \left( \frac{\dot  R}{R}\right)^2-\frac{\ddot R}{R}\right]=\frac{1}{2}\left( \frac{1}{\Omega_{\Lambda}}-1 \right)~~~,\cr
\end{align}
with $\Omega_{\Lambda}$ the dark energy fraction.
The perturbation equations of Eq. \eqref{eq:finalpert2} now become
\begin{align}\label{eq:finalpert3}
A^{(1)\prime\prime}=&\frac{3}{4}\frac{A^9}{z^4B^4}+\frac{3 \xi A^5}{4 z^4}~~~,\cr
B^{(1)\prime\prime}=&-\frac{9}{4}\frac{A^8}{z^4B^3}+\frac{15 \xi A^4 B} {4 z^4}-\frac{3 \eta A^5}{z^4}~~~,\cr
\end{align}
which can be integrated to give corrections to the zeroth order solutions for use as initial values in
solving the $A$ and $B$ differential equations.

\end{document}